\documentclass[10pt,aps,pre,amsmath,amsfonts,amssymb,floatfix,longbibliography,superscriptaddress,notitlepage, onecolumn]{revtex4-1}
\usepackage{mathrsfs} \usepackage{graphicx,color} \usepackage{bbm} \usepackage{multirow}
\usepackage[normalem]{ulem}
\usepackage{lineno}
 \let\t=\tau

\newcommand{\red}[1]{\textcolor{black}{#1}}

\newcommand{\blue}[1]{\textcolor{black}{#1}}
\newcommand{\green}[1]{\textcolor{black}{#1}}
\newcommand{\modi}[1]{\textcolor{black}{#1}}
\newcommand{\magenta}[1]{\textcolor{black}{#1}}
\newcommand{\add}[1]{\textcolor{black}{#1}}

\def\to{\rightarrow}

\newcommand{\eq}[1]{Eq.~(\ref{#1})}
\newcommand{\beq}{\begin{equation}} \newcommand{\eeq}{\end{equation}}

\renewcommand{\phi}{\varphi}
\renewcommand{\vec}[1]{\boldsymbol{#1}}

\begin{document}

%\title{Anomalous response of amorphous solids under shear} 
 
\title{Exploring the complex free energy landscape of the simplest glass by rheology}

\author{Yuliang Jin}
\email{jinyuliang@cp.cmc.osaka-u.ac.jp}

\affiliation{Cybermedia Center, Osaka University, Toyonaka, Osaka 560-0043, Japan}

\author{Hajime Yoshino}
\affiliation{Cybermedia Center, Osaka University, Toyonaka, Osaka 560-0043, Japan}
\affiliation{Graduate School of Science, Osaka University, Toyonaka, Osaka 560-0043, Japan}

\begin{abstract}
  \textbf{
        \magenta{For amorphous solids, it has been intensely debated whether the traditional view on solids,  in terms of the ground state and harmonic low energy excitations on top of it\green{, such as phonons,} is still valid.}
 %has been a subject of intense debates.}
      %\blue{For amorphous solids, the traditional view on solids that they can be described well in terms of the ground state and harmonic low energy excitations on top of it, such as phonons,
 %has been a subject of intense debates.}
   %\modi{
    %\sout{The essence of the traditional view on solids is that they can be described in terms of the ground state and harmonic low energy excitations on top of it, such as phonons.} }
    %While 
    %\modi{it} is \modi{well} 
    %\modi{the harmonic description (phonons) is well} established for crystals, %the validity of the simple harmonic picture \modi{(phonons)} 
    %\modi{its validity} 
    %for amorphous solids has been a subject of intense debates\modi{.}
  %  \blue{For amorphous solids,}  \sout{The essence of} the traditional view on solids \sout{is} that they can be described \blue{well} in terms of the ground state and harmonic low energy excitations on top of it, such as phonons
    %\sout{. While this \modi{simple harmonic picture is well} established for crystals, \modi{its} validity 
    %of the simple harmonic picture 
    %for amorphous solids} has been a subject of intense debates\modi{.}
    %\cite{MS86,ML99,combe2000strain,pratt2003nonlinear,schuh2003nanoindentation,maloney2006amorphous,HKEP11,rodney2011modeling,YO12,YZ2014PRE,OH14,muller2015marginal,denisov2016universality,BU2016NP,NYZ2015}.
    %Recent 
    %\modi{\sout{emerging}} 
    \blue{Recent  theoretical developments} \green{of amorphous solids}
    %\sout{of a class of first-principle \modi{theories}  of amorphous solids}
    %~\cite{CKPUZ14,RUYZ2015PRL}
    revealed the possibility of %\modi{an} 
    unexpectedly
    complex free energy \modi{landscapes where the 
    %traditional 
   simple 
    harmonic picture breaks down.}
   % is predicted to break down\modi{.}
     %\cite{YZ2014PRE,BU2016NP}.
 Here we demonstrate that 
 %\modi{\sout{a}} 
 \modi{standard rheological techniques} can be used as 
 %\modi{\sout{a}} 
 \modi{powerful tools to 
 %\sout{explore} 
 examine non-trivial consequences of such complex free energy landscapes.} 
 %\sout{the} complex free energy landscape\sout{ and examine the intriguing theoretical predictions}.
 By 
 %\modi{\sout{performing}} 
 extensive numerical simulations on a hard sphere glass under quasi-static shear at finite temperatures,
we show that, above the so-called Gardner transition density, the elasticity breaks down, the stress relaxation exhibits slow and aging dynamics, and the apparent shear modulus becomes protocol-dependent. Being designed to be reproducible in \modi{laboratories}, our approach may 
%\sout{open the door 
%of opportunity 
%to explore} 
\blue{trigger explorations of} the complex free energy landscapes of a large variety of amorphous materials. 
}
%Maximum is 150 words without references.
\end{abstract}

\maketitle

\modi{\section*{Introduction}}

Amorphous and crystalline solids have very different behaviors under external perturbations, \modi{especially \blue{rheological properties under} shear deformations~\cite{MS86,ML99,combe2000strain,pratt2003nonlinear,schuh2003nanoindentation,maloney2006amorphous,HKEP11,rodney2011modeling,YO12,YZ2014PRE,OH14,muller2015marginal,NYZ2015,BU2016NP,denisov2016universality,BU2016NP,PRSS2016PRE,franz2016mean,OY13}.}
%\modi{\cite{,OY13}}. 
It is well known that by increasing the shear strain, a crystal displays a linear elastic response, followed by plastic deformation and yielding. 
%\modi{In the elastic regime, the crystal  can be described in terms of the ground state and harmonic low energy excitations on top of it, such as phonons. 
\modi{However,}
experiments and numerical simulations show that this picture breaks down for amorphous solids, such as glasses~\modi{\cite{HKEP11,OY13,rodney2011modeling,PRSS2016PRE,ML99,dubey2016elasticity,schuh2003nanoindentation,maloney2006amorphous,NYZ2015}},
%\cite{,ML99,dubey2016elasticity,schuh2003nanoindentation,maloney2006amorphous,NYZ2015}, 
granular matter~\cite{denisov2016universality,combe2000strain,OH14}, and foams~\cite{pratt2003nonlinear}, where the elastic behavior is mixed with plastic events. Such plastic events cause sudden drops in stress-strain curves, and are sometimes referred to as crackling noise~\cite{sethna2001crackling}, due to their similarities to avalanches in earthquakes. An apparent shear modulus or \add{ rigidity} $\mu$, which is the ratio between the stress and strain, can be nevertheless defined and measured. Experiments on glassy emulsion systems~\cite{mason1995elasticity, mason1997osmotic} show that $\mu$ scales linearly $\mu \sim P$ with the pressure $P$ both below and above the jamming density,
while  harmonic treatments predict $\mu \sim {P^{1.5}}$ (below)~\cite{brito2006rigidity}
and $\mu \sim {P^{0.5}}$ (above)~\cite{OSLN03} respectively.
%below and above the jammming density respectively.
These contradictions reveal that amorphous solids can be strikingly softer than purely harmonic solids like crystals, even at sufficiently low temperatures where the harmonic expansion was conventionally expected to be valid.

On the theoretical side, the mean-field  theory based on the exact solution in the large dimensional limit of the hard sphere glass has brought a more accurate and comprehensive picture beyond the harmonic description~\modi{\blue{\cite{KPZ12,KPUZ13,CKPUZ13,YZ2014PRE,CKPUZ14, RUYZ2015PRL,rainone2016following,BU2016NP}}}. The main outcome is the prediction of a Gardner transition 
%density $\varphi_{\rm G}$  
(see Fig.~\ref{fig:yield}a), which divides the classical amorphous phase into two: in the \add{ stable phase} (or \add{ normal phase}), the state is confined in one of the simple smooth basins on the free energy landscape; once the system is compressed above the Gardner transition density $\varphi_{\rm G}$ (or is 
cooled down below the  Gardner transition temperature $T_{\rm G}$), the simple glass basin splits into  a fractal hierarchy of sub-basins and the glass state becomes marginally stable. \modi{Although similar ideas of complex energy landscapes have been conceived phenomenologically in earlier works \blue{(see \cite{heuer2008exploring} and references there in)}, 
%However, being a first principle theory, 
the mean-field theory gives a firmer first principle ground for such a picture, 
%and provides 
with falsifiable predictions. 
In particular, the theory predicts that the elastic anomalies and non-trivial rheology should only appear in the \add{ marginally stable phase} (or \add{ Gardner phase})~\cite{YZ2014PRE,BU2016NP}, 
%suggesting non-trivial rheology
which lies deep inside the glassy phase.
However, the mean-field theory is exact only in the large dimensional limit, and its relevance 
    in real systems is far from obvious.
Here we test the theoretical proposal \blue{of the non-trivial rheology} in physically relevant dimensions $d=2$ and 3, and compare quantitatively the theoretical predictions
%\blue{scaling laws $\to$ theory}  
with our numerical data. }

\modi{We design laboratory reproducible \blue{rheological protocols 
to examine the signatures of the intriguing complex free-energy landscape.}
%\sout{\modi{based on standard rheological schemes}}, 
%and apply them 
Our protocols are applied on densely packed hard spheres, which is a simple and representative glass-forming model. Our result shows the anticipated anomalous rheology emerging at the Gardner transition, which turns out to be strikingly similar to the dynamical responses of spin glasses to an external magnetic field~\cite{nordblad1998experiments,vincent2007ageing}. The evidence of a complex free energy landscape in 
the Gardner phase is consistent with a previous numerical study~\cite{BCJPSZ2016PNAS}, where particles' vibrational dynamics is analyzed. 
%The approach in~\cite{BCJPSZ2016PNAS}
That approach
 has been used in a  recent experiment of an agitated granular system~\cite{seguin2016experimental}. 
%By analyzing particles' vibrational dynamics, a recent experiment has shown evidences of the Gardner transition in an agitated granular system~\cite{seguin2016experimental}. 
However, generalizing the method to other systems, such as molecular glasses, may not be easy due to the difficulty of tracking  trajectories of individual particles.
The approach proposed in this study overcomes this problem, since it requires no microscopic information, but only the standard macroscopic rheological measurements (the shear stress and strain) that are well accessible in many experimental systems, including molecular and metallic glasses, polymers and colloids.}
In the present \modi{paper}, however, we do not attempt to judge whether the Gardner transition survives in finite dimensional systems as a sharp phase transition or becomes a
  crossover (in the thermodynamic limit),
but rather we aim to explore the possibilities to observe its non-trivial signatures
in experimentally feasible length/time scales.
%\modi{\sout{To avoid crystallization, we work on a polydisperse mixture of hard spheres  whose diameters are distributed according to a probability distribution $P(D) \sim D^{-3}$, for $D_{\rm min} \leq D < D_{\rm min}/0.45$.
%~\cite{BCNO2016PRL,BCJPSZ2016PNAS}. 
%The Gardner transition of this model has been detected by characterizing the growing time and length scales associated to the thermal vibrations of particles
%~\cite{BCJPSZ2016PNAS}
%.} }

%\paragraph*{Breakdown of elasticity --} 

\modi{\section*{Results}}

\modi{{\bf Preparation of stable glasses.}} \modi{To avoid crystallization, we work on a polydisperse mixture of hard spheres  whose diameters are distributed according to a probability distribution $P(D) \sim D^{-3}$, for $D_{\rm min} \leq D < D_{\rm min}/0.45$~\cite{BCNO2016PRL,BCJPSZ2016PNAS} (see Supplementary Note 1)}.
A glass is typically obtained by a slow compression (or cooling) annealing from a dilute state, where it falls out of equilibrium at the compression (or cooling) rate-dependent glass transition density  $\varphi_{\rm g}$  (or glass transition temperature $T_{\rm g}$). Since we choose hard spheres as our working system, the density is the control parameter. 
%To mimic this glass forming process,  

We design a numerical protocol to mimic a simple shear experiment of deeply annealed glasses (see Fig.~\ref{fig:yield}).  Our protocol includes three steps: 
%(i)
We first use the swap algorithm~\cite{BCNO2016PRL,BCJPSZ2016PNAS} to prepare a well-equilibrated\modi{, supercooled-liquid} configuration at various densities $\varphi_{\rm g}$ \modi{(see Supplementary Methods and \add{Supplementary Figure} 1)}. The algorithm combines the Lubachevsky-Stillinger algorithm~\cite{SDST06}, which consists of standard event-driven molecular dynamics (MD) and slow compression, with  Monte-Carlo swaps of particle diameters.
\modi{The MD time is expressed in units of $\sqrt{\beta m \bar{D}^2}$,
  where the particle mass $m$ and mean diameter $\bar{D}$, as well as the inverse temperature $\beta$, are all set to unity.
  \blue{In other words, a particle travels over a distance 
    %the hardspheres travel over distances 
    of the order of the diameter
    within a unit MD time. }}
%, to prepare a well-equilibrated configuration at various densities $\varphi_{\rm g}$ (see SI).
%At a given density $\varphi_{\rm g}$, we prepare many of such equilibrated configurations, which are statistically independent from each other, and we call them  {\it samples} in the following.
From the thermodynamic point of view, the system is still in the liquid
%  state as long as $\varphi_{\rm g}$ is smaller than the putative
%  Kauzmann density $\varphi_{\rm K}$
but we work at
density $\varphi_{\rm g}$ sufficiently above the mode-coupling theory (MCT) crossover density $\varphi_{\rm d}$.
%{\color{green} FC: do we need to introduce $\varphi_{\rm K}$?}
Then once we switch off the particle swapping 
  and return to the natural dynamics simulated by MD,
  the $\alpha$-relaxation time \modi{has become}  much larger than our \modi{MD} simulation time scales so that the system behaves essentially as a solid. 
  This glass is thus ultrastable, in a sense similar to those obtained by 
  %in silico 
  vapour deposition experiments~\cite{PRRR14,LQMJH14,YTGER15}.
At a given density $\varphi_{\rm g}$, we prepare many of such equilibrated configurations, which are statistically independent from each other, and we call them  \add{ samples} in the following.  
  %yu2015suppression}.
  
  %(ii) 
\add{ Second, subsequently} the equilibrated configuration is compressed up to a target density $\varphi$
%,  using the Lubachevsky-Stillinger (LS) algorithm, that combines EDMD and a constant growth of particle 
  with a compression rate $\delta_{\rm g} = 10^{-3}$. 
  %\blue{[Define the time unit.]}
From a single sample, that is a starting equilibrated configuration at $\varphi_{\rm g}$, we generate an ensemble of
  compressed glasses at $\varphi$, obtained by choosing statistically independent initial particle velocities 
  %of the particles
  drawn from the Maxwell-Boltzmann distribution. 
  We call each of such compressed glasses as a \add{ realization} in the following.
  \modi{These realizations are out-of equilibrium, since they no longer follow the liquid equation of state (EOS), but we consider that they remain in
    \add{ restricted equilibrium}~\cite{RUYZ2015PRL} for $\varphi < \varphi_{\rm G}$, i.e., they are equilibrated within the given glass state determined by the sample. The MD  preserves the kinetic energy
    so that the system remains at the unit temperature throughout our simulations. \blue{The typical scale of the vibrations
      of the particles within the glass states depends on $\varphi$.
      For instance it varies from
      $10^{-1}$ to $10^{-2}$ for 
      $\varphi=0.645$ to $\varphi=0.688$ (for $\phi_{\rm g}=0.643$, see Fig.~2 of \cite{BCJPSZ2016PNAS}), so that
      particles make $10$ to $10^{2}$ collisions within a unit MD time.}
  }

%(iii) 
\add{ Third, for} a given realization,  a simple shear is applied.  The simple shear  is modelled by an affine deformation of the $x$-coordinates of all particles, $x_i \to x_i + \gamma z_i$, under the Lees-Edwards boundary condition~\modi{\cite{lees1972computer}} with fixed system volume. The shear strain is increased quasi-statically with \add{ a small constant shear rate $\dot{\gamma} = 10^{-4}$, such that the shear rate dependence is negligible in the regime $\varphi < \varphi_{\rm G}$} 
\modi{(see \add{Supplementary Figure}~6 for a discussion on the $\dot{\gamma}$-dependence)},
%(see SI for a discussion on the $\dot{\gamma}$-dependence), 
and the shear stress $\Sigma$
  is measured at different $\gamma$. The shear stress $\Sigma$ and the pressure $P$ are both calculated from inter-particle interactions due to collisions between hard sphere particles. %\modi{\sout{(see SI)}}.
 For convenience, we introduce reduced pressure $p = \beta P/\rho$ and reduced stress $\sigma=\beta \Sigma /\rho$,
    where 
   % \modi{\sout{$\beta=1/(k_{\rm B}T)$ (set to unity) is the inverse temperature and}}
    $\rho$ is the number density of the particles
(see \modi{Supplementary Note 1}). 
  Note that as the pressure, the shear stress is entirely due to momentum exchanges between the particles
  so that the rigidity is purely entropic in hard sphere systems.
 \modi{Furthermore, because shear stress and pressure has the same physical dimension,
  it is convenient to introduce a  rescaled stress $\tilde{\sigma}=\sigma/p$.}
  
%an instantaneous increment  $\delta \gamma = 10^{-4}$ at each step, followed by a number of MD collisions such that the shear rate is $\dot{\gamma} = 10^{-4}$.

{\bf Breakdown of elasticity.}
Figure~\ref{fig:yield} shows the phase diagram for our polydisperse hard sphere model, and typical stress-strain curves of individual realizations in different density regimes. 
%(i) %At the equilibrium density $\varphi = \varphi_{\rm g}$
In the stable glass phase $\varphi_{\rm g} < \varphi < \varphi_{\rm G}$ (Fig.~\ref{fig:yield}b),
% where $\varphi_{\rm G}$ is the Gardner transition density,  
the stress-strain curve shows a smooth linear (harmonic) response regime at small $\gamma$, followed by a
%sudden 
sharp drop of the stress $\sigma$, signalling the yielding of the system. At yielding, a system-wide shear band emerges  (see Fig.~\ref{fig:yield}c), and the system is driven out of a free energy metastable glass basin. 
%After yielding, it \red{becomes a fluid, such that different basins are explored freely}. 
\modi{After yielding, the system enters a steady flow state, similar to those observed in athermal amorphous solids under quasi-static shear~\cite{maloney2006amorphous, karmakar2010statisticalB}.}
%(ii) 
In the Gardner phase $\varphi_{\rm G} < \varphi < \varphi_{\rm J}$, \modi{where $\varphi_{\rm J}$ is the jamming density,}
%\sout{the elastic regime shrinks to zero}
the harmonic response is punctuated by
%\blue{Did we check? see \cite{ML99} Fig 14. which demonstrates this. Subsequently Procaccia's group says the same thing later \cite{HKEP11}. See the comment below.},
 \modi{\add{ mesoscopic plastic events} (MPEs)} that can happen at  very small $\gamma$ (see Fig.~\ref{fig:yield}d). These \modi{MPEs} correspond to sudden avalanche-like heterogeneous rearrangements of particle positions without formation of band-like patterns (see Fig.~\ref{fig:yield}e). 
% The jerky nature of the stress-strain curve reveals that 
%the metastable free energy glass basin splits into many sub-basins, and the barrier-crossing between these sub-basins results in the \modi{mesoscopic plastic event}.
%Note that similar 
\modi{Similar MPEs} have been observed in quasi-static shear simulations 
  at zero temperatures \cite{ML99,maloney2006amorphous},
but our simulations are performed at finite temperatures. \modi{Note that 
%here we show stress-strain curves of a single realization of the compressed glass to display the breakdown of elasticity. The 
the details of the plastic events, including the locations of yielding, jamming, and MPEs, depend on the samples \modi{(see \add{Supplementary Figure}~8)} and realizations \modi{(see \add{Supplementary Figure}~7)}. 
%See \modi{Supplementary Notes 2 and 3, \add{Supplementary Figure}s~2, 3, and 4} 
For the behavior  of the stress-strain curves averaged over many realizations and samples, see \add{Supplementary Figures}~2, 3, and 4, as well as Supplementary Notes 2 and 3.}

%%\blue{In the \cite{ML99} the break down of linear elasticity is seen.  See Fig 1 and Fig2. Also you can see this in Maloney-Lemaitre's paper. 
%They could observe these because they are working at zero temperature.
%Your simulation is instead done at finite temperatures so that these could be seen only in the Gardner's phase.  May be we can comment on this in the text?}
%\red{In the Malando-Lack's paper, Fig 11 is like our Fig. 1 e)}.

For large $\varphi$,  the stress $\sigma$ grows dramatically at large $\gamma$, and appears to diverge (see Fig.~\ref{fig:yield}d).
%, and yielding becomes impossible. 
This \add{ shear jamming} phenomenon is due to the dilatancy effect of hard sphere glasses under shear: the pressure $p$ increases with $\gamma$ when the system volume is fixed.  Note that if $p$ is kept as a constant when $\gamma$ is increased, then the volume expands due to the dilatancy effect. In that case, shear jamming does not appear and shear yielding is recovered (see \modi{\add{Supplementary Figure}~5}).  While the switching from shear yielding to shear jamming with increasing $\varphi$ is not a consequence of the Gardner transition, it implies that the system is trapped more deeply in the metastable basin, and that the activated barrier-crossing between metastable basins becomes forbidden. However, the emergence of sub-basins in the Gardner phase implies that even though the usual relaxation ($\alpha$-relaxation) is frozen, an additional slow dynamics may appear. This aspect is explored below.

%(i) In the stable glass phase $\varphi_{\rm g} < \varphi < \varphi_{\rm G}$  (Fig.~\ref{fig:yield}c), the smooth linear response regime persists for small $\gamma$. The stress overshoots the linear response before yielding occurs -- a phenomenon becomes more profound at higher densities (see Fig.~\ref{fig:yield}d). This overshooting shows that, compared to the equilibrium case in (i),  the out-of-equilibrium system is trapped deeper in the free energy metastable glass basin, and yielding, which corresponds to barrier-crossing over basins, becomes more difficult. 
%In the large $\gamma$ regime, the stress-strain curve is not smooth -- some sudden and small drops of stress are observed, which correspond to avalanche events of the rearrangement of particle positions. The separation between the smooth linear and the jerky regimes is a signature of Gardner transition under shear. 
%{\bf Discuss the connection to the \modi{mesoscopic plastic event} in athermal granular systems.}

{\bf Aging and slow dynamics.} We next show that in the Gardner phase,  the relaxation of shear stress becomes complicated, accompanied by aging and a slow dynamics.
\modi{Due to the similarity between the Gardner transition and the spin glass transition, here it is very useful to firstly recall what happens in spin glasses,  which are essentially  disordered and highly frustrated magnets
  \cite{BY86,mydosh1993spin}.
  The mean-field spin glass theory
   %which is exact in the large dimensional limit 
   has suggested complex free energy landscapes of spin glasses manifested as continuous replica symmetry breaking~\cite{parisi1987spin}, much as what happens
  in the Gardner phase of hard sphere glasses~\cite{CKPUZ13,CKPUZ14}.
  Remarkably, this feature is \blue{predicted to have a reflection in the dynamics},  resulting in non-trivial dynamical 
  responses to external magnetic field, and aging effects in the relaxation of magnetization~\cite{cugliandolo1993analytical,cugliandolo1994out,franz1998measuring}. In experiments, the simplest approach to examine the intriguing features of the dynamics is a combination of the so called zero-field cooling (zfc) and field cooling (fc) protocols.
  In the zfc protocol, one cools a spin glass sample from a high temperature in the paramagnetic phase down to a target temperature $T$, where a magnetic field $h$ is switched on and one 
  %starts to 
  measures the increase of the magnetization. In the fc protocol, one first switches on the
magnetic field $h$, and then subsequently cools the system down to the target temperature $T$ and measures the remanent magnetization.
The key point is that, in the two protocols, the order of cooling and switching on of the magnetic field is \add{ reversed}.
In such experiments~\cite{nagata1979low,aruga1994experimental},
the magnetizations observed in the zfc/fc protocols are the
same if the working temperature $T$ is higher than the spin glass transition temperature, 
%$T_{\rm g}$,
while the fc magnetization  becomes larger than the zfc magnetization if $T$ is lower than the spin glass transition temperature.
%while the magnetization of the fc protocol becomes larger than that of the zfc protocol if $T$ is lower than the spin glass transition temperature.
%$T_{\rm g}$.
The anomaly, i.~e., the difference between the zfc and fc magnetizations,  is naturally explained by the mean-field theory
\cite{parisi1987spin}.
Furthermore, examinations of the aging effects by these protocols give detailed information about the complex free energy landscape~\cite{nordblad1998experiments,vincent2007ageing,cugliandolo1993analytical,cugliandolo1994out,franz1998measuring}.
}

\modi{It has been pointed out theoretically that the shear on structural glasses plays a very similar role as the magnetic field on spin glasses~\cite{YM10,YO12}, and that the relaxation of the shear stress should also reflect the complex free energy landscape
  encoded by the continuous replica symmetry breaking solution in the Gardner phase~(see \modi{\add{Supplementary Figure} 9}, and Fig.~2 of \cite{YZ2014PRE}). The shear strain and stress in structural glasses correspond to the magnetic field  and  magnetization in spin glasses respectively. Furthermore, 
 apparently compression corresponds to cooling in the hard sphere glasses.
  Therefore,} inspired by the 
 % zero-field cooling/field cooling 
  \modi{zfc/fc}
  experiments in spin glasses, we \modi{design} two distinct protocols which are combinations of compression and shear  exerted in reversed orders 
% \modi{\sout{(see Fig.~\ref{fig:illustration})}}
: In the \add{ zero-field compression} (ZFC) protocol, we first compress the configuration from $\varphi_{\rm g}$ to $\varphi$, \red{and set the time to zero}. 
We then wait for time $t_{\rm w}$ before a shear strain $\gamma$ is applied instantaneously (see \modi{Supplementary Methods}), and 
% \sout{(after reseting time to zero)} 
measure the relaxation of the stress $\sigma_{\rm ZFC} (t, t_{\rm w})$ \red{as a function of the time $\tau=t-t_{\rm w}$ elapsed after switching on the strain}.
%A longer waiting time $t_{\rm w}$ in the aging protocol is equivalent to a slower compression rate $\delta_{\rm g}$ in the continuous compression protocol.
On the other hand, in the \add{ field compression} (FC) protocol, we first apply an instantaneous increment of shear strain at the initial density $\varphi_{\rm g}$,  compress the configuration to $\varphi$, and set the time to zero.
%\sout{, wait for $t_{\rm w}$,}
Then we measure the relaxation of the stress \red{$\sigma_{\rm FC}(t)$
  as the function of the elapsed time $t$. 
 }

For $\varphi < \varphi_{\rm G}$, 
\modi{no aging effect is observed, and the dynamics is fast. The}
%\sout{both} 
 $\sigma_{\rm ZFC} (t, t_{\rm w})$
%\sout{and $\sigma_{\rm FC} (t, t_{\rm w})$} 
\red{is stationary or time translationally invariant (TTI), i. e.
  $\sigma_{\rm ZFC} (\tau, t_{\rm w})=\sigma_{\rm ZFC} (\tau)$,
  depending only on the time difference $\tau=t-t_{\rm w}$ but not on the waiting time $t_{\rm w}$  (see Fig.~\ref{fig:dynamics}).}
 % as expected for equilibrium systems (see Fig.~\ref{fig:dynamics}).}
%independent of the waiting time $t_{\rm w}$
After a time scale {\modi{$\tau_{\rm b}$} 
%\blue{\sout{$\sim O(1)$} [
 %     ...it seems shorter than $O(1)$...
  %    the unit MD time corresponds to the time needed for particles to travel over distance $\bar{D}$, which is much greater than cage sizes...]}
    corresponding to the ballistic motions of particles~\cite{BCJPSZ2016PNAS}, the ZFC stress  $\sigma_{\rm ZFC} (\tau, t_{\rm w})$ converges quickly  to $\sigma_{\rm FC}(t)$ which is almost a constant in time. 

In contrast, for $\varphi > \varphi_{\rm G}$, $\sigma_{\rm ZFC} (\tau, t_{\rm w})$
%\sout{and $\sigma_{\rm FC} (t, t_{\rm w})$} 
displays strong $t_{\rm w}$-dependent aging effects \red{manifesting the out-of-equilibrium nature of the system}, \modi{as well as a slow  dynamics.}
% such that these quantities explicitely depends
%on two times. 
In such a situation, different large time limits can emerge
depending on the order of $\tau \to \infty$ and $t_{\rm w} \to \infty$ \cite{bouchaud1998out}.
An important feature which can be seen in Fig.~\ref{fig:dynamics}
  is that $\sigma_{\rm ZFC}(\tau,t_{\rm w})$ exhibits
  a plateau suggesting the existence of a large time limit
  $\sigma_{\rm ZFC} \equiv  \lim_{\tau \to \infty}\lim_{t_{\rm w} \to \infty} \sigma_{\rm ZFC} (\tau, t_{\rm w})$ where $t_{\rm w} \to \infty$ is taken \add{ before}
  $\tau \to \infty$.
    On the other hand $\sigma_{\rm FC}(t)$ is again essentially constant in time $t$ (for \modi{$t > \tau_{\rm b}$}) 
    %\sout{for not too small $t_{\rm w}$}, 
    and we shall denote it as $\sigma_{\rm FC}$. In the reversed order of the large time limits, we expect that the ZFC shear stress decays to the FC one, $\lim_{t_{\rm w} \to \infty}\lim_{\tau \to \infty} \sigma_{\rm ZFC} (\tau, t_{\rm w})=\sigma_{\rm FC}$. However, the convergence becomes slower as  $t_{\rm w}$ increases, and its corresponding time scale could be beyond the simulation time window, as shown in the case of Fig.~\ref{fig:dynamics}.

Apparently $\sigma_{\rm ZFC}$ is larger than $\sigma_{\rm FC}$ when $\varphi > \varphi_{\rm G}$, which implies
the ergodicity breaking.
The aging effect and the slowing down of dynamics show the similarities between the Gardner transition and the liquid-glass transition, which demonstrates that the Gardner transition could be considered as a \blue{``glass transition within the glass phase''} \blue{(see also Supplementary Note 3)}.
%the 2nd section
% entitled as ``Relaxation of the shear stress: connection to the free-energy landscape''.)}
%These dynamical effects are secondary, to be distinguished from those occurring around the glass transition. 
In a sharp contrast, because the Gardner transition is absent in a crystal, its shear stress relaxes faster when $\varphi$ increases, \modi{and no aging is present}.

{\bf Protocol-dependent shear modulus.} The above observation suggests that the linear shear moduli measured by the two protocols should be distinct in the Gardner phase. We determine the \blue{apparent} shear modulus $\mu$
%\blue{\sout{from the linear relations} 
\modi{as} $\mu_{\rm ZFC} = (\sigma_{\rm ZFC}- \sigma_0)/\gamma$ and $\mu_{\rm FC} = (\sigma_{\rm FC}- \sigma_0)/\gamma$,  where $\sigma_0$ is the remanent shear stress at $\varphi$ before $\gamma$ is applied. The shear strain is increased quasi-statically with rate $\dot{\gamma} = 10^{-4}$ up to 
a predetermined small \blue{target} $\gamma$.
% = 2 \times 10^{-3}$. 
The shear stress is measured at $\tau=1$ after waiting for  $t_{\rm w} = 10$.
%, which corresponds to the plateau values $\sigma_{\rm ZFC} (t \sim \tau_{\rm cage}, t_{\rm w} \to \infty)$ and $\sigma_{\rm FC} (t \sim \tau_{\rm cage}, t_{\rm w} \to \infty)$ in Fig.~\ref{fig:dynamics} 
%(however, note the difference that data are obtained for one single sample in Fig.~\ref{fig:dynamics}, but averaged over many samples in Fig.~\ref{fig:modulus}). 
Details on the \modi{time and $\varphi_{\rm g}$ dependences} of the shear modulus is discussed in \modi{\add{Supplementary Figures 10,} 13, and 14}.

Figure~\ref{fig:modulus} shows that, while $\mu_{\rm ZFC}$ and $\mu_{\rm FC}$ are indistinguishable in the stable glass phase $\varphi < \varphi_{\rm G}$ (or $p < p_{\rm G}$), they become clearly distinct in the Gardner phase $\varphi > \varphi_{\rm G}$ (or $p > p_{\rm G}$). ~\modi{ For a similar  result of a two-dimensional bidisperse hard disk model, see \add{Supplementary Figure} 16.}
%This basic phenomenon
%-- the separation of  $\mu_{\rm ZFC}$ and $\mu_{\rm FC}$ at $\varphi_{\rm G}$ -- 
%is independent of system size $N$  and the glass transition density $\varphi_{\rm g}$ -- we observe that  $\mu_{\rm ZFC}$ and $\mu_{\rm FC}$ start to bifurcate around $\varphi_{\rm G}$,  for all the systems considered in this study.  
\modi{This behabior of shear modulus is a consequence of the time dynamics of the shear stress illustrated in Fig.~\ref{fig:dynamics}: at the time scales used to measure the shear modulus ($\tau = 1$ and $t_{\rm w} = 10$), the two shear stresses $\sigma_{\rm ZFC}$ and $\sigma_{\rm FC}$ have converged to the same value for $\varphi<\varphi_{\rm G}$, but remain different for $\varphi > \varphi_{\rm G}$.}
The bifurcation point determines the Gardner transition threshold $\varphi_{\rm G}$ (or $p_{\rm G}$). Within the numerical accuracy, the $\varphi_{\rm G}$ determined from this approach is fully consistent with the previous estimate based on particles' vibrational motions and caging order parameters~\cite{BCJPSZ2016PNAS}.  To further test this result, we perform detailed analysis on its dependence on the number of particles $N$ and the shear strain $\gamma$,
% and the initial equilibrium density $\varphi_{\rm g}$, 
as discussed below.

We find no appreciable finite size effects for $\mu_{\rm FC}$ (see Fig.~\ref{fig:modulus}a), which is in contrast to the observation in non-equilibrated systems, where $\mu_{\rm FC}$ decreases to zero in the thermodynamic limit~\cite{NYZ2015}. It suggests that preparing deeply equilibrium configurations is the key
to observe the \modi{non-vanishing $\mu_{\rm FC}$}. 
%, which is generally challenging, is crucial in order to reproduce similar phenomena in other systems. 
\modi{While the shear moduli measured around the Gardner transition, and therefore the determination of $\varphi_{\rm G}$, are $N$-independent, stronger finite size effects are observed for $\mu_{\rm ZFC}$ at large $p$ near the jamming limit:
%We expect that in the ZFC protocol, compared to smaller systems,  larger systems have more stress relaxation channels (i.e., more frequent \modi{mesoscopic plastic event} events) under shear. 
%\modi{Interestingly, we observe that smaller systems obey better the mean-field theory. Indeed, in Ref.~\cite{karmakar2010statistical}, the authors measured numerically the mean strain $\delta \gamma$ at which the first \modi{mesoscopic plastic event} event takes place in amorphous solids, and found a finite-size scaling
%$\delta \gamma \sim N^{\beta}$ with $\beta \approx -0.62$. The scaling suggests that, in larger systems, \modi{mesoscopic plastic event} is easier to occur, which reduces the measured $\mu_{\rm ZFC}$ (see Fig.~\ref{fig:yield}) to a value that departs further away from the mean-field theoretical prediction (recall that the mean-field $\mu_{\rm ZFC}$ does not consider any \modi{mesoscopic plastic event} contribution).  As $N \to \infty$, \modi{mesoscopic plastic event} becomes unavoidable at any finite shear strain since $\delta \gamma \to 0$.}
$\mu_{\rm ZFC}$ is lower in larger systems, suggesting a stronger non-linear effect.}
%(see \modi{Supplementary Note 4}).}
Nevertheless, the data of $\mu_{\rm ZFC} (p)$\modi{, with a fixed $\gamma$,} appear to converge for $N \gtrsim 2000$, which confirms that $\mu_{\rm ZFC} (p)$ and  $\mu_{\rm FC} (p)$ remain distinguishable in the thermodynamic limit, for $\varphi > \varphi_{\rm G}$. 
%, and more frequent avalanche events. 

Regarding the $\gamma$-dependence, Fig.~\ref{fig:modulus}b shows that,  within the numerical accuracy,  $\mu_{\rm FC}$ is independent of $\gamma$, as long as $\gamma$ is sufficiently small. On the other hand, for $\varphi > \varphi_{\rm G}$ \modi{and a given $N$}, $\mu_{\rm ZFC}$ slightly increases with decreasing $\gamma$. This result shows that in the Gardner phase, the non-linear effect on  $\mu_{\rm ZFC}$  remains even for very small $\gamma$, which is consistent with the observation of elasticity breakdown in Fig.~\ref{fig:yield}. \modi{Such non-linear effects are observed for any $N$ studied (see \modi{\add{Supplementary Figure} 12}), and we expect that in the thermodynamic limit $N \to \infty$, a pure linear behavior of $\mu_{\rm ZFC}$ can only exist in the limit $\gamma \to 0$~\cite{karmakar2010statistical}. The vanishing of the pure elastic regime distinguishes the Gardner phase from the normal glass and crystalline phases.  For a more detailed discussion on how the shear moduli depend on the strain $\gamma$, the particle number $N$, the initial density $\varphi_{\rm g}$, and the waiting time $t_{\rm w}$, see Supplementary Note 4.}

%We next compare our simulation data with theoretical and experimental results. For $\varphi < \varphi_{\rm g}$, we compare our data to the mean-field {\it state following} theory~\cite{RUYZ2015PRL} by plotting rescaled $\mu/\mu_{\rm g}$ as a function of $p/p_{\rm g}$, where $\mu_{\rm g}$ and $p_{\rm g}$ are the shear modulus and the pressure at $\varphi_{\rm g}$.  On this rescaled plot, the theory and simulation data show similar behaviors, both of which are insensitive to $\varphi_{\rm g}$. 
%Note that the mean-field theory uses an over-simplified liquid equation of state, that is only valid for mono-disperse hard spheres in the large dimensional limit. Thus a direct comparison between the theory and our simulation is impossible. However, our results show that the theory captures the general trend on how the system evolves under a slow compression annealing, which resembles the state following procedure in the theory.

For $\varphi > \varphi_{\rm g}$, 
%theoretical results of neither $\mu_{\rm ZFC}(p)$ nor $\mu_{\rm FC}(p)$ have been reported. However, 
the mean-field theory predicts two power-law scalings in the large $p$ limit~\cite{YZ2014PRE}: $\mu_{\rm ZFC} \sim p^{\kappa}$ \modi{with $\kappa=1.41574...$}, and  $\mu_{\rm FC} \sim p$. \modi{The first scaling has also been derived semi-empirically by an independent approach \cite{degiuli2014force}.}
%Comparing our numerical data to the mean-field theoretical prediction (see Fig.~\ref{fig:modulus}a and 3b), 
We find good agreement  between the theory and simulation on the scaling of $\mu_{\rm FC}$ (see Fig.~\ref{fig:modulus}). For $\mu_{\rm ZFC}$, a noticeable discrepancy is observed in the limit of large $N$ for a fixed finite $\gamma$ (Fig.~\ref{fig:modulus}a), but the discrepancy decreases when $\gamma \to 0$ for a fixed $N$ (Fig.~\ref{fig:modulus}b),  \modi{or when $N$ is decreased for a fixed $\gamma$ (Fig.~\ref{fig:modulus}a). This is because the mean-field $\mu_{\rm ZFC}$ is obtained in the pure linear response limit \blue{$\gamma \to 0$}, while the non-linear effect caused by MPEs would increase  with $\gamma$ and $N$,  as discussed above.} 
%\red{Here should we cite Procaccia's scaling $\Delta \gamma_1 \sim N^{-0.62}$, where $\Delta \gamma_1$ is the strain at which the first \modi{mesoscopic plastic event} happens?}
The scaling  $\mu_{\rm FC} \sim p$ is consistent with the experimental observation in emulsions ~\cite{mason1995elasticity, mason1997osmotic}. Considering the experimental system is possibly not deeply equilibrated, we expect that the relaxation of experimental $\mu(t)$ is sufficiently fast, and the measurement was performed in the long time limit $\mu(t\to\infty ) \to \mu_{\rm FC}$ (see the discussion of Fig.~\ref{fig:dynamics}).

%We observe that the numerical data of  $\mu_{\rm ZFC}$ agree better with the theoretical scaling for smaller $N$ and smaller $\gamma$. The result suggests that the numerical model is closer to a mean-field-like system in this limit, due to the lack of \modi{mesoscopic plastic event} relaxations. 

%The numerical power-law exponents are $\kappa_{\rm ZFC} = 1.20(1)$ and  $\kappa_{\rm FC} = 1.00(2)$  for $N=8000$ and $\varphi_{\rm g}=0.643$. The two exponents appear to be $\varphi_{\rm g}$-independent within the available accuracy. Compared to the mean-field theoretical prediction,  $\kappa_{\rm ZFC}^{\rm mf} =1.41574$  and $\kappa_{\rm FC}^{\rm mf} = 1.0$~\cite{YZ2014PRE}, we find a full agreement on $\kappa_{\rm FC}$ but a discrepancy on $\kappa_{\rm ZFC}$. This discrepancy might be caused by finite dimensional corrections. 

%To further verify and emphasize the protocol dependence on the shear modulus, we design a third protocol, which applies an additional shear strain after the FC procedure. The shear modulus measured by the third protocol is close to $\mu_{\rm ZFC}$, and clearly different from $\mu_{\rm FC}$~(Fig.~\ref{fig:modulus}d).
%{\color{green} FC: this paragraph could go into SI.}

{\bf Interpretation of results.}
%To understand the reason why $\mu_{\rm FC} < \mu_{\rm ZFC}$ in the Gardner phase, 
The Gardner transition is a consequence of the split of glass basins in the phase space~\cite{CKPUZ14}, and the split of particle cages in the real space (see Fig.~\ref{fig:illustration}). The schematic plot of the free energy $F$ as a function of $\gamma$ in Fig.~\ref{fig:illustration} illustrates how a glass basin splits into many sub-basins once the system is compressed above $\varphi_{\rm G}$. Here we interpret our results based on this free energy landscape viewpoint. First,  in the ZFC protocol, the system intends to remain in one of the sub-basins after compression \modi{(note that different realizations may end up in different sub-basins)}, but as $\gamma$ increases in a quasi-static shear procedure (Fig.~\ref{fig:yield}), it may become unstable where the shear stress drops abruptly, resulting in a \modi{MPE}. The \modi{MPE} could be interpreted as shear-induced barrier crossing between sub-basins, analogous to the barrier crossing between basins in a yielding event. Second,  if $\gamma$ is fixed, the shear stress relaxes with time, and 
%the relaxation time scale grows exponentially with the activation barrier, 
according to the Arrhenius law, the emergence of barriers between sub-basins would result in a slowing down of the relaxation dynamics  with $\varphi$ (Fig.~\ref{fig:dynamics}).
%The slowing down of the relaxation dynamics  with $\varphi$ (Fig.~\ref{fig:dynamics}) is therefore consistent with a growth of the barriers between sub-basins. 
The appearance of aging further reveals the emergence of complex structures within a basin, similar to the mechanism of aging in the glass transition \cite{bouchaud1998out}. Third, because in the FC protocol, the system can overcome the sub-basin barriers, the $\mu_{\rm FC}$ always corresponds to the second order curvature of basins, rather than that of sub-basins as in the $\mu_{\rm ZFC}$ case. This results in $\mu_{\rm FC} < \mu_{\rm ZFC}$ for the regime $\varphi > \varphi_{\rm G}$ as observed in Fig.~\ref{fig:modulus}.
\modi{Note that according to Fig.~\ref{fig:illustration}e, one should obtain a shear modulus close to $\mu_{\rm ZFC}$ if an additional strain is applied after FC, as confirmed in \add{Supplementary Figure}~15.}
On the other hand, no basin split occurs and therefore two protocols are equivalent  in the stable phase $\varphi < \varphi_{\rm G}$. Previous study~\cite{NYZ2015} has shown that $\mu_{\rm ZFC} = \mu_{\rm FC}$ for crystals. The similarity between crystals and stable glasses further confirms that their free energy basins are similarly structureless.

\modi{\section*{Discussion}}
 \modi{We wish to stress that,}
our data cannot exclude that the Gardner transition becomes just a crossover in finite dimensions, such that no real phase transition exists.
% (e.g., this is certainly true in two dimensions {\color{yellow} ref}). 
Yet, irrespective of the sharpness of the Gardner transition, we rationalize here in a unified framework all the observations obtained on the rheological behavior of the simple hard sphere glass, and find quantitatively reasonable agreement between the theory and simulations. 
Thus, even if the Gardner transition is not sharp in the thermodynamic limit, 
 for accessible sizes in numerical simulations, and likely for those in experiments as well~\cite{seguin2016experimental}, a behavior reminiscent of the transition can be clearly observed.

 Finally, we make remarks on experimental consequences.
\red{It is an intriguing question to clarify whether the phase diagram presented in Fig.~\ref{fig:yield}a is generic
 in a wide range of amorphous solids, ranging from different kinds of glasses to
  soft matter such as colloids (One can choose to change the temperature or pressure as
 the control parameter depending on specific systems).}
 %\modi{\sout{By analyzing particles' vibrational dynamics, a recent experiment has shown evidences of the Gardner transition in an agitated granular system~\cite{seguin2016experimental}. However, generalizing the method to other systems, such as molecular glasses, may not be easy due to the difficulty to track the trajectories of individual particles.
% Our proposed ZFC/FC approach (Fig.~\ref{fig:illustration}) overcomes this problem, since it requires no microscopic information, but only the standard macroscopic rheological measurements that are well accessible in many experimental systems.}}
\red{The crucial point is to keep track of the dynamical effects which might have been overlooked in some previous experiments, for the following two reasons. First, 
in reasonably stabilized dense systems,
 the liquid EOS (green line in Fig.~\ref{fig:yield}a) and
 the Gardner line (red line) becomes separated enough, so that
 the liquid dynamics ($\alpha$-relaxation) and  the intriguing
 internal  glassy dynamics ($\beta$-relaxation  induced by the Gardner transition) can be well separated in time scales.
In this respect,
 % The remaining key challenge is to prepare long aged glasses, analogous to those obtained by swap algorithm.
recently developed experimental techniques, such as the vapor deposition~\cite{PRRR14,LQMJH14,YTGER15} and the high pressure path~\cite{paluch2003does}, or the use of sufficiently old natural glasses~\cite{zhao2013using},
 %sufficiently old on a geological time scale
   would provide ideal settings.
If such an ideal setting is not possible, 
%then the liquid EOS and the Gardner line is not so well separated (see the vicinity of $\varphi_{\rm d}$ in Fig.~\ref{fig:yield}a), and different dynamical time scales are mixed. In these cases, 
one could freeze the $\alpha$-relaxation out of the experimental time window, by working 
at sufficiently low temperatures or high densities.
%so that $\alpha$-relaxation is excluded out of the experimental time window.
The second reason is that by experimentally studying the aging effects due to the
 internal dynamics of the amorphous solids, the complexity of the
 free energy landscape could become manifested as we demonstrated in the present \modi{paper}.} \\

%\modi{{\bf Data availability.} The data that support the  findings of this study are available in Osaka University Knowledge Archive (OUKA) with the identifier [DOI will be provided].}\\

%The data that support the findings of this study are available from the corresponding author upon request.} \\

%\modi{{\bf Author contributions.} All authors contributed to all aspects of this work.} \\

%\modi{{\bf Competing financial interests.} The authors declare no competing financial interests.} \\

\acknowledgments

%We especially thank Francesco Zamponi, Daijyu Nakayama, and Satoshi Okamura for many useful discussions.
\modi{We thank Giulio Biroli, Ludovic Berthier, Patrick Charbonneau, Olivier Dauchot, Ana\"el Lema\^{\i}tre, Corrado Rainone, and Pierfrancesco Urbani for useful discussions, and especially Francesco Zamponi, Daijyu Nakayama, and Satoshi Okamura for many stimulating interactions. }
This work was supported by KAKENHI (No. 25103005  ``Fluctuation \& Structure'') from MEXT, Japan. The Computations were performed using Research Center for Computational Science, Okazaki, Japan, \blue{and the computing facilities in the Cybermedia center, Osaka Univerisity.}

\clearpage

\begin{figure*}
\centerline{\includegraphics[width=\columnwidth]{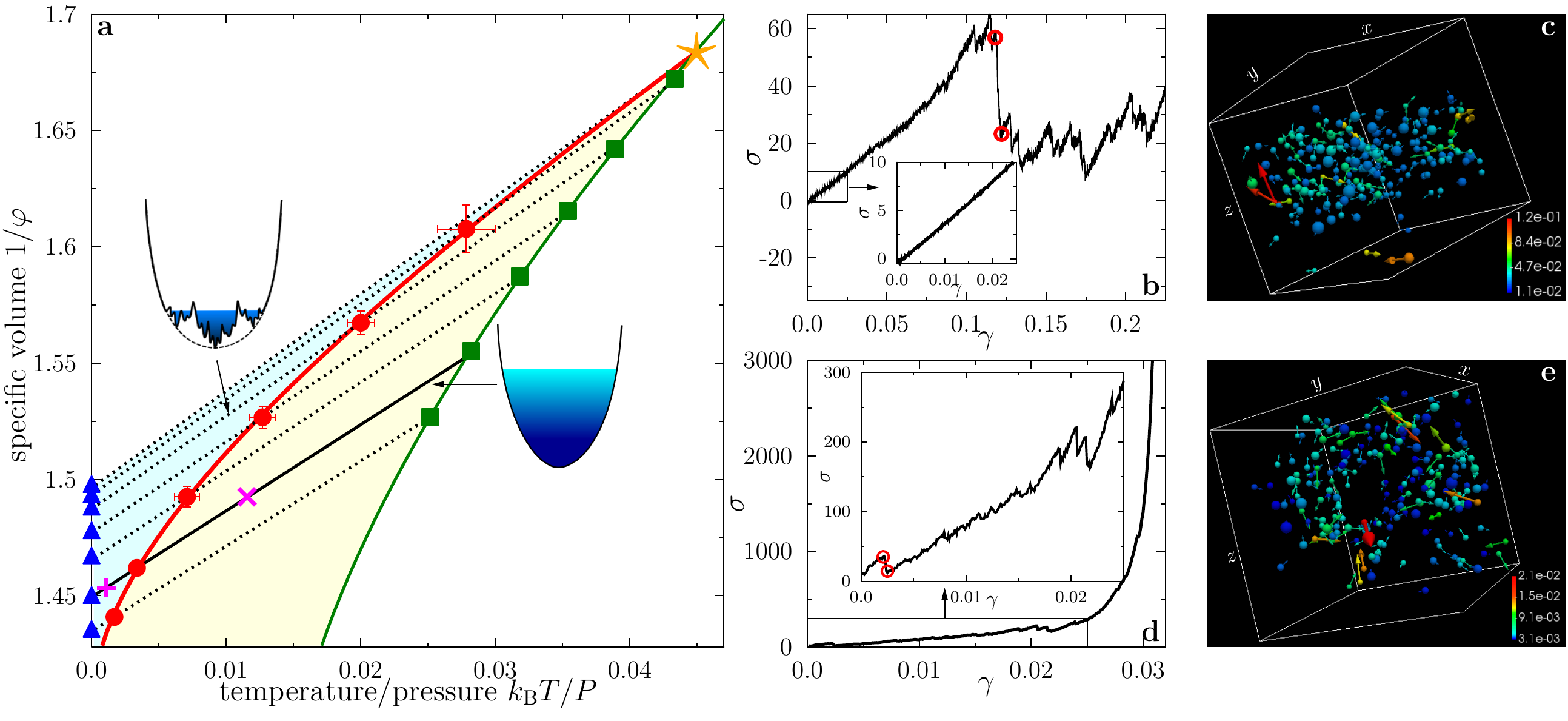} }
\caption{{\bf Typical stress responses under quasi-static shear.} (a) Illustration of the protocol on the polydisperse hard sphere glass phase diagram (adapted from Ref.~\cite{BCJPSZ2016PNAS}), where $k_{\rm B}T/P = 1/(\rho p)$. The MCT dynamical crossover (yellow star) is located at $\varphi_{\rm d} = 0.594(1)$ along the equilibrium liquid \modi{EOS} (green line).
  %with $\rho$ and $p$ being the particle number density and the reduced pressure respectively.
Using the swap algorithm we first prepare equilibrium samples
  at various densities $\varphi_{\rm g}$  (green squares)
whose pressure obeys the \modi{Carnahan-Stirling empirical liquid EOS \blue{(solid green line)} ~\cite{BCJPSZ2016PNAS}}. 
%\blue{[Determined numerically or an empirical formula?]}
%  The equilibrium \blue{samples obtained} by the swap algorithm at $\varphi_{\rm g}$  (green squares) obeys the liquid equation of state (EOS, green line).
Next we switch off the swap algorithm, and
   perform compression annealing from $\varphi_{\rm g}$ to jamming (blue triangles),  producing realizations of compressed glasses at various densities $\varphi$.
   The system is now out-of-equilibrium and the pressure follows the glass EOSs  \blue{$p \propto 1/(\varphi_{\rm J}-\varphi)$} (black dotted lines) \blue{~\cite{BCJPSZ2016PNAS}}. The Gardner transition $\varphi_{\rm G}$ (red circles and line) separates the stable (light yellow regime) and the marginally stable  (light blue regime) glass phases. The insets show schematic depictions of free energy landscapes in these two different phases.
  As an example,  an equilibrium configuration is prepared at $\varphi_{\rm g}=0.643$, and compressed (solid black line) up to $\varphi_{\rm J} = 0.690(1)$.
  We show typical stress-strain  curves under quasi-static shear \blue{with increasing $\gamma$}, \red{using a single realization of the compressed glass of $N=1000$ particles}, at (b) $\varphi = 0.670$ (pink cross) and (d) $\varphi = 0.688$ (pink plus), which are below and above $\varphi_{\rm G} = 0.684(1)$ respectively.
Curves in (b) and (d) are zoomed in (insets) for  $\gamma \leq 0.025$, to show the different small-$\gamma$ behaviors in the two cases.
The real-space vector fields of particle displacements are visualized in (c) for a yielding event (\blue{between the two} red circles in (b)),  and (e) for a \blue{mesoscopic plastic event (\modi{MPE})} (\blue{between the two} red circles in (d)), where each sphere
%\sout{locates} 
\blue{is located} at the equilibrium position before yielding/\modi{MPE}, and each  
vector represents the displacement during yielding/\modi{MPE}. We have subtracted the affine part caused by shear from the displacements, and only show top $20\%$ particles with large displacements. A shear band around the middle of the $z$-axis is observed in (c).  The sizes of particles are reduced by a factor of 0.4, and the vectors are amplified in length by a factor of 2 in (c) and a factor of 15 in (e). The color represents the magnitude of displacement. }
\label{fig:yield}
\end{figure*}
\clearpage

\begin{figure}[h]
\centerline{\includegraphics[width=0.6\columnwidth]{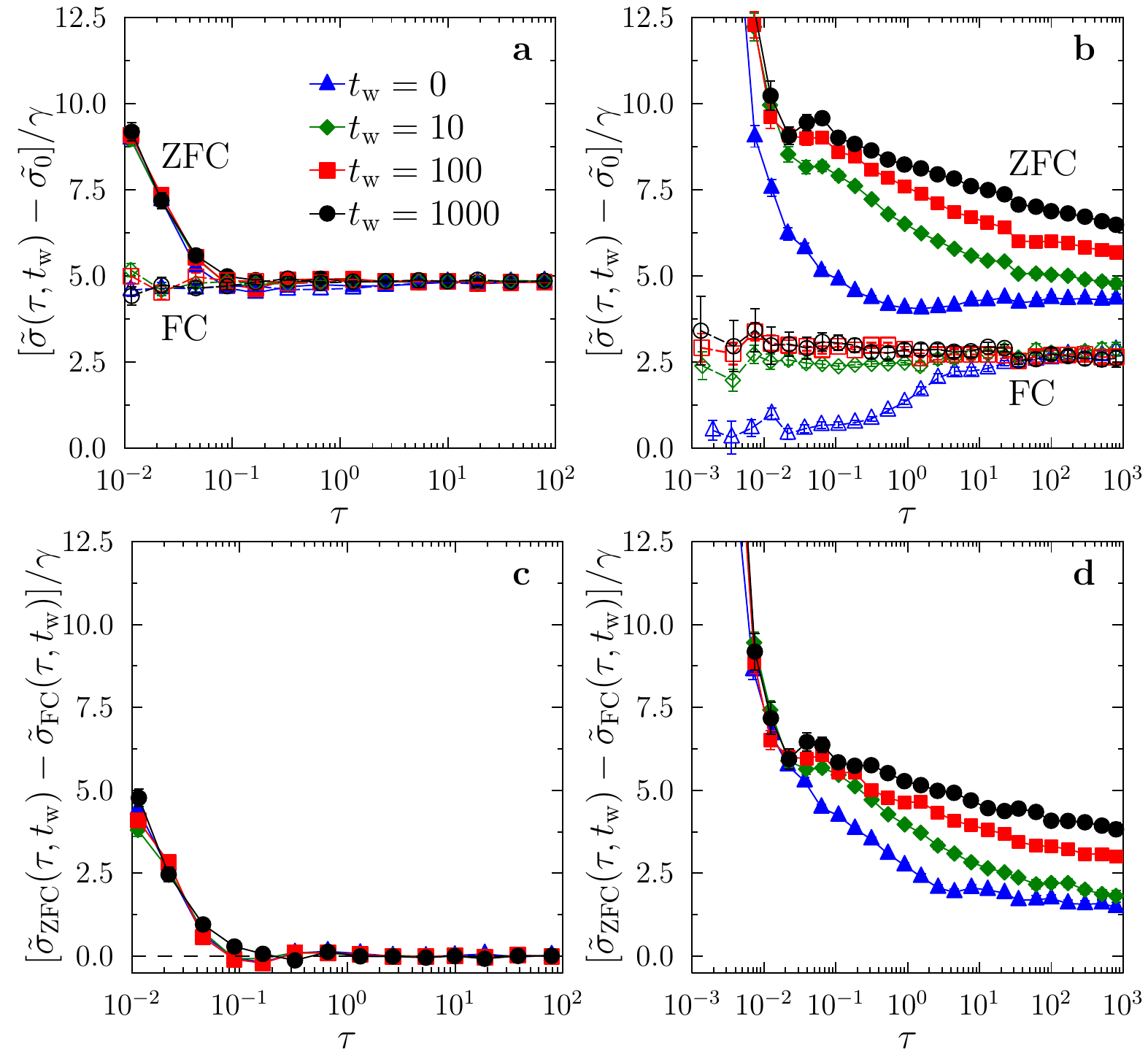} }
\caption{{\bf Relaxation of shear stress.}
%  \blue{[$\tilde{\sigma} \to \sigma$, because $\sigma$ is defined as the reduced shear stress $\sigma=\Sigma/p$.]}
  Relaxations of the rescaled ZFC shear stress $\tilde{\sigma}_{\rm ZFC} = \sigma_{\rm ZFC}/p$ (filled symbols) and the rescaled FC shear stress $\tilde{\sigma}_{\rm FC} = \sigma_{\rm FC}/p$ (open symbols) show different behaviors at (a) $\varphi  = 0.670$ and (b) $\varphi = 0.688$, corresponding to the pink plus and cross in Fig.~\ref{fig:yield} respectively \modi{(the Gardner transition density $\varphi_{\rm G} = 0.684(1)$~\cite{BCJPSZ2016PNAS})}. We show results for several different waiting time $t_{\rm w}$, under an instantaneous increment of shear strain $\gamma = 10^{-3}$. Data are averaged over many realizations of compressed glasses
  obtained from a single equilibrated sample at $\varphi_{\rm g}=0.643$
  with $N=1000$ particles. Here 
%the rescaled shear stress $\tilde{\sigma} = \sigma/p$, and $\tilde{\sigma}_0$ is 
the rescaled remanent stress $\tilde{\sigma}_0$ is measured in the ZFC protocol at $\varphi$, after the longest waiting time $t_{\rm w} = 1000$ and before the shear strain is applied. The difference $\tilde{\sigma}_{\rm ZFC} (\tau, t_{\rm w}) - \tilde{\sigma}_{\rm FC}  (\tau, t_{\rm w}) $ quickly vanishes  and does not show significant  $t_{\rm w}$-dependence at (c) $\varphi = 0.670$, while it decays much slower and shows a strong $t_{\rm w}$-dependent aging effect at (d) $\varphi = 0.688$. Note that by definition, $\tilde{\sigma}_{\rm FC} (t)$ is a one variable function, but we plot it here as $\tilde{\sigma}_{\rm FC} (\tau, t_{\rm w})$ in order to compare it with $\tilde{\sigma}_{\rm ZFC} (\tau, t_{\rm w})$. \modi{The pressure $p$ is independent of time and protocol, in both cases (see \add{Supplementary Figure~11}). The error bars denote the standard error of the mean (s.e.m.).}
% \blue{Pressure: which pressure is used here? the pressure depends on time?}}
}
\label{fig:dynamics}
\end{figure}
\clearpage

\begin{figure}
\centerline{\includegraphics[width=0.7\columnwidth]{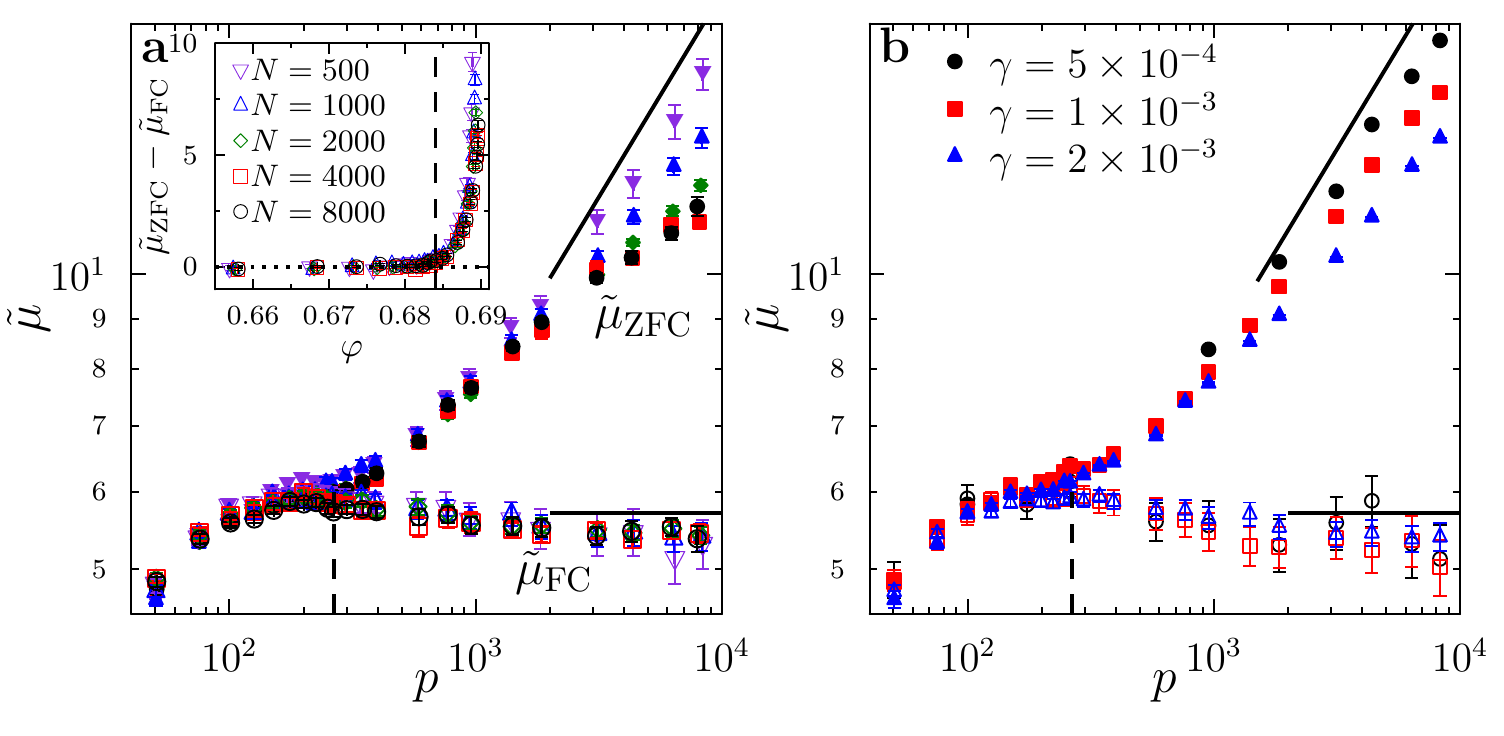} }
\caption{{\bf Protocol-dependent shear modulus.} (a) The rescaled shear modulus $\tilde{\mu} =  \mu/p$,
  %\red{[May be we can avoide introducing this and write simply as $\mu/p$. Introduction of too many 'reduced quantities' can be confusing.]}
  obtained from both ZFC (filled symbols)  and FC (open symbols), is plotted as a function of $p$, for $\varphi_{\rm g} = 0.643$ and several $N$. 
  %The value of $\mu$ is obtained from $(\sigma - \sigma_0)/\gamma$ with $\gamma = 2 \times 10^{-3}$. 
  The data are obtained by using  $\gamma = 2 \times 10^{-3}$, and are averaged over 
%$N_{\rm s}  = 50- 250$ samples, and $N_{\rm r} = 20- 100$ 
$N_{\rm s} \approx 200$ samples, and $N_{\rm r} \approx 100$ 
individual realizations for each sample. The two shear moduli $\mu_{\rm ZFC}$ and $\mu_{\rm FC}$ coincide below $p_{\rm G}$ (vertical dashed line), and become distinct above, %Best fittings (black solid lines) to the large-pressure data of $N=8000$ give $\kappa_{\rm ZFC} =1.20(1)$ and $\kappa_{\rm FC} = 1.00(2)$, which
where  \modi{$p_{\rm G}=265$~\cite{BCJPSZ2016PNAS}}.
The data are compared to the large $p$ scalings predicted by the mean-field theory $\mu_{\rm ZFC} \sim p^{1.41574}$ and  $\mu_{\rm FC} \sim p$ (black solid lines). 
The difference $\mu_{\rm ZFC} - \mu_{\rm FC}$ is plotted as a function of $\phi$ in the inset\modi{, where the vertical dashed line represents $\varphi_{\rm G} = 0.684$~\cite{BCJPSZ2016PNAS}}. 
(b) Rescaled ZFC and FC shear moduli obtained from  a few different $\gamma$, for $N=1000$ systems. \modi{The error bars denote the s.e.m.}
%(c) The numerical data of a few different $\varphi_{\rm g}$ are compared to the mean-field theory (lines) with $\hat{\varphi}_{\rm g} = 5,6,7,8$, where the theoretical $\hat{\varphi}_{\rm g} = 2^d \varphi_{\rm g} /d$ in $d$ dimensions, following the convention used in Ref.~\cite{RUYZ2015PRL}. Both numerical and theoretical results are rescaled by the reference values $\tilde{\mu}_{\rm g}$ and $p_{\rm g}$ at $\varphi_{\rm g}$. The inset shows the difference $\mu_{\rm ZFC} - \mu_{\rm FC}$ as a function of $\phi$ for a few different $\varphi_{\rm g}$, where the Gardner transitions $\varphi_{\rm G}$ (values from Ref.~\cite{BCJPSZ2016PNAS}) are marked by vertical lines. (d) The reduced shear modulus obtained from the third protocol (FC+shear) is compared to $\tilde{\mu}_{\rm ZFC}$ and $\tilde{\mu}_{\rm FC}$, for  $\varphi_{\rm g} = 0.643$, $N=1000$ and $\gamma = 2 \times 10^{-3}$. {\color{green} FC: maybe move panels (c) and (d) to SI.}
%The theoretical $\varphi_{\rm g}$ is expressed in reduced unit $\hat{\varphi} = 2^d \varphi /d$. 
}
\label{fig:modulus}
\end{figure}
\clearpage

\begin{figure}
\centerline{\includegraphics[width=0.6\columnwidth]{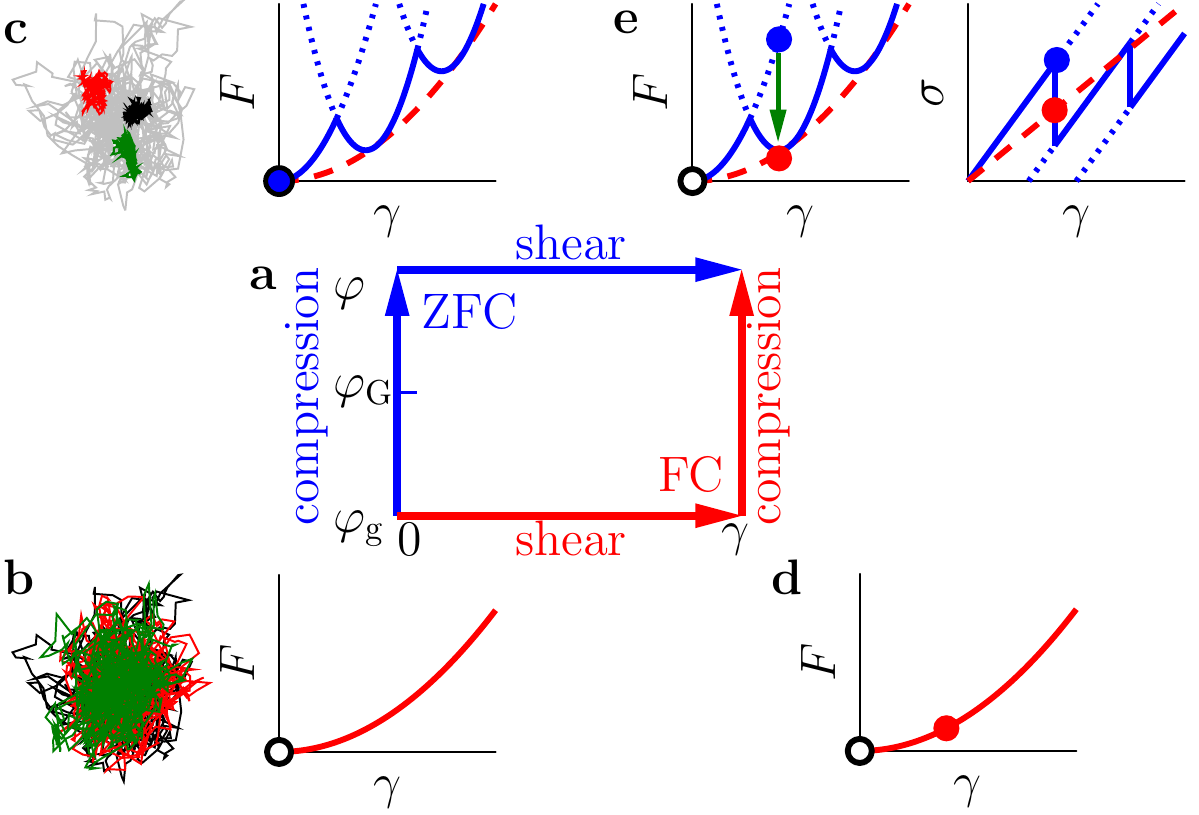} }
%\caption{{\bf Interpretation of results within a free energy landscape picture.}
\caption{ 
     \add{ {\bf Illustration of protocols.}  We show the evolution of the free energy landscape and the state point  $(\varphi, \gamma)$
      under compression and shear.} \modi{(a)} In the ZFC protocol, the system is first compressed and then sheared, while
      the order is reversed in the FC protocol.
     \red{
    %(center)
%    The illustration of the ZFC and the FC protocols.
    %(bottom-left)      
       \modi{(b) State point $(\varphi_{\rm g}, 0)$:}
       The schematic free energy $F$ as a function of the strain $\gamma$ at the initial density $\varphi = \varphi_{\rm g}$ before compression. We assume that the initial state point (black open circle) is located at the minimum of the parabola.
       To show an example of  the real-space particle caging, we also plot three independent trajectories of the same tagged particle
       in the same two-dimensional sample
       %in a two-dimensional system
       (see \modi{Supplementary Note 5}).
       % as an example. 
       %(top-left) 
       \modi{(c) State point $(\varphi, 0)$:} If the system is compressed first to $\varphi$ (above the Gardner transition density $\varphi_{\rm G}$), the free energy basin (red dashed line) splits into many sub-basins (blue line):
       %and the particle's cage splits into many small ones
        the state point (blue solid circle) becomes trapped in one of the sub-basins.
       The dotted blue lines represent the metastable region of the sub-basins. The split of free energy basin corresponds to the split of cage in the real space \modi{(as an example, see the independent trajectories representing three split cages)}. 
       %(bottom-right) 
       \modi{(d) State point $(\varphi_{\rm g}, \gamma)$:} On the other hand, if the system is sheared first, the state point (red solid circle) is forced to climb up the parabola of the basin. 
%       How the state is perturbed in the free energy basin by shear without compression (red solid circle).
       \modi{(e) State point $(\varphi, \gamma)$:}
       %(top-right) 
       After both shear and compression, the state point can be located at different points
       in the same free energy landscape, depending on the order of the compression and shear. In the ZFC case,
       the state point (blue solid circle) is forced to climb up the sub-basin where it is trapped,
       while it can remain at lower energy state  
       in the FC protocol (red solid circle). Because sub-basins are meta-stable (dotted  blue line), \modi{MPEs occur} with increasing $\gamma$ in a quasi-static shear, and slow relaxation occurs for a fixed $\gamma$ (green arrow). The shear stress $\sigma$ is determined by $\sigma \sim dF/d\gamma$~\modi{(right panel)}, and the  shear modulus  by $\mu = d \sigma/d\gamma \sim d^2F/d\gamma^2$. The stress-strain curves show that for $\varphi > \varphi_{\rm G}$, $\mu_{\rm ZFC}$ (slope of blue line) is larger than $\mu_{\rm FC}$ (slope of dashed red line).}
%We plot schematically the free energy $F$ as a function of $\gamma$ at the initial density $\varphi = \varphi_{\rm g}$ before compression (left), and that at a target density $\varphi >\varphi_{\rm G}$ after compression (right). 
%In (a) ZFC, the system is perturbed within sub-basins (blue line), while in (b) FC, the barriers between sub-basins are overcome during compression, which results in a lower shear modulus. The red line represents the basin re-organized from sub-basins, which is the envelope of the blue line. Because sub-basins are meta-stable (dotted  blue line), \modi{mesoscopic plastic event} occurs with increasing $\gamma$ in a quasi-static shear, and slow relaxation occurs for a fixed $\gamma$ (dashed green arrow). Insets are three independent  trajectories of the same particle in a two-dimensional system (see SI), which shows how the particle cage splits in the real-space. 
}
%This illustration shows that  the  distinction of $\mu_{\rm FC}$ and $\mu_{\rm ZFC}$ is a consequence of the split of metastable basin in the Gardner phase.}
\label{fig:illustration}
\end{figure}

\clearpage

\appendix
\setcounter{figure}{0}  
\renewcommand\theequation{Supplementary Equation \arabic{equation}}
\renewcommand{\figurename}{Supplementary Figure}

{\centerline{\LARGE {\bf Supplementary Information}}}
\bigskip
\bigskip
\bigskip
\bigskip
\bigskip
\bigskip

\begin{figure}[h]
\centerline{\hbox{\includegraphics[width=0.3\columnwidth]{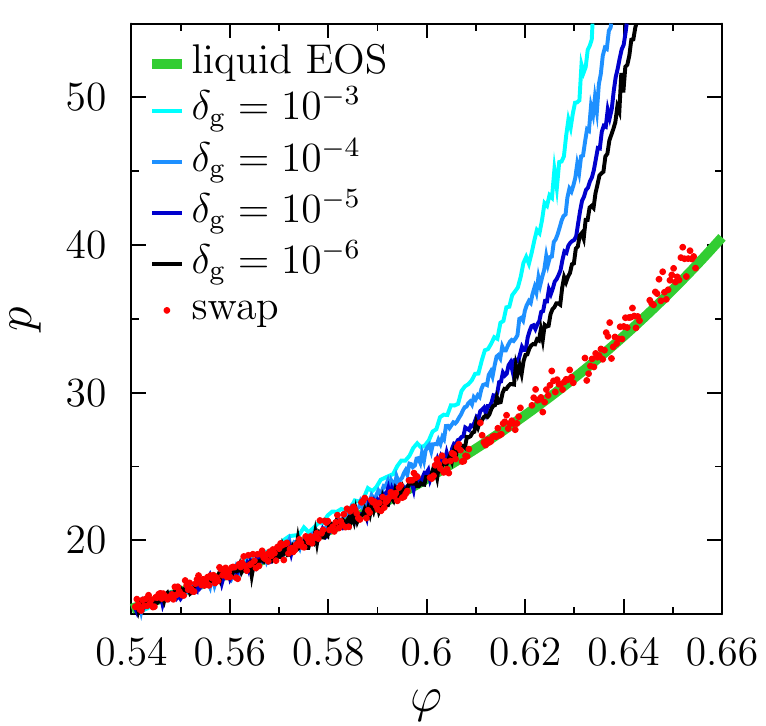}}}
\caption{Evolution of the reduced pressure $p$ under compression, for a  system of $N=2000$ particles. The data obtained from the swap algorithm agree with the \modi{Carnahan-Stirling empirical liquid EOS \blue{\eq{eq:poly_EOS}} ~\cite{BCJPSZ2016PNAS}}.
%equilibrium liquid EOS. \blue{[Determined numerically or an empirical formula?]} 
For comparison, we also plot data obtained from pure compression done by the LS algorithm without the swap for a few different compression rate $\delta_{\rm g}$. The swap algorithm  falls out of equilibrium at much higher $\varphi$, compared to the standard compressions.}
\label{fig:swap}
\end{figure}

\begin{figure}[h]
\centerline{\hbox{\includegraphics[width=0.6\columnwidth]{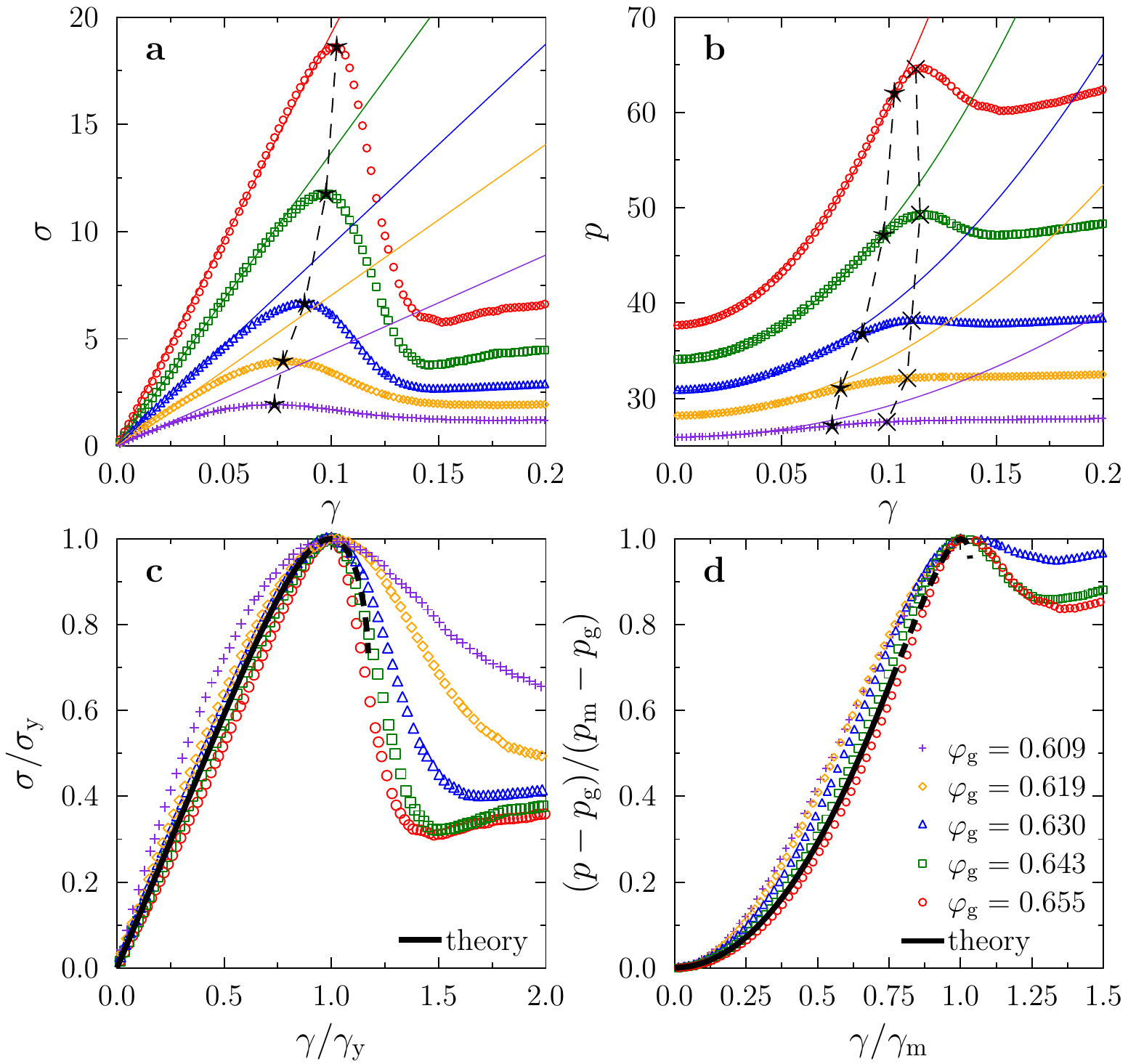}}}
\caption{Quasi-static shear on equilibrium configurations prepared at a few different $\varphi_{\rm g}$.  The systems consist of $N=1000$ particles, and the data are averaged over $N_{\rm s} = 100$ samples and $N_{\rm th} = 10-20$ realizations for each sample. (a) The shear stress $\sigma$ and  (b) the pressure $p$  are plotted as functions of the shear strain $\gamma$. At small $\gamma$, the stress-strain curve is fitted to a linear function $\sigma = \mu \gamma$ (lines), and the pressure-strain curve is fitted to a quadratic function $p = p_{\rm g} + R \gamma^2$ (lines).
%, where $p_{\rm g}$ is the reduced pressure at $\varphi_{\rm g}$ and $\gamma =0$, and $R$ is the dilatancy parameter.  
  The star marks the peak of the stress-strain curve, which represents the yielding point ($\gamma_{\rm y}, \sigma_{\rm y}$), and the cross marks the peak ($\gamma_{\rm m}, p_{\rm m}$) of the pressure-strain curve.  The parameters $\mu$, $\gamma_{\rm y}$, $\sigma_{\rm y}$, $\gamma_{\rm m}$, $p_{\rm m}$, and $R$ are reported in Supplementary Figure~\ref{fig:parameters} as functions of $\varphi_{\rm g}$.  (c) The rescaled stress-strain curves and (d) the rescaled pressure-strain curves are compared to the mean-field theoretical predictions (black line)~\cite{RUYZ2015PRL},
\red{for the equilibrium volume fraction $\hat{\varphi}_{\rm g} =  2^d \varphi_{\rm g} /d = 7 $, \blue{where the dimension $d=3$}. Here the solid line part is the stable 1-step replica symmetry breaking (1RSB) solution, and the dashed line part is the unstable 1RSB solution~\cite{RUYZ2015PRL}. We have checked that the theoretical results are insensitive to $\hat{\varphi}_{\rm g}$ on these rescaled plots}.
}
\label{fig:SF_phig}
\end{figure}

\begin{figure}[h]
\centerline{\hbox{\includegraphics[width=0.6\columnwidth]{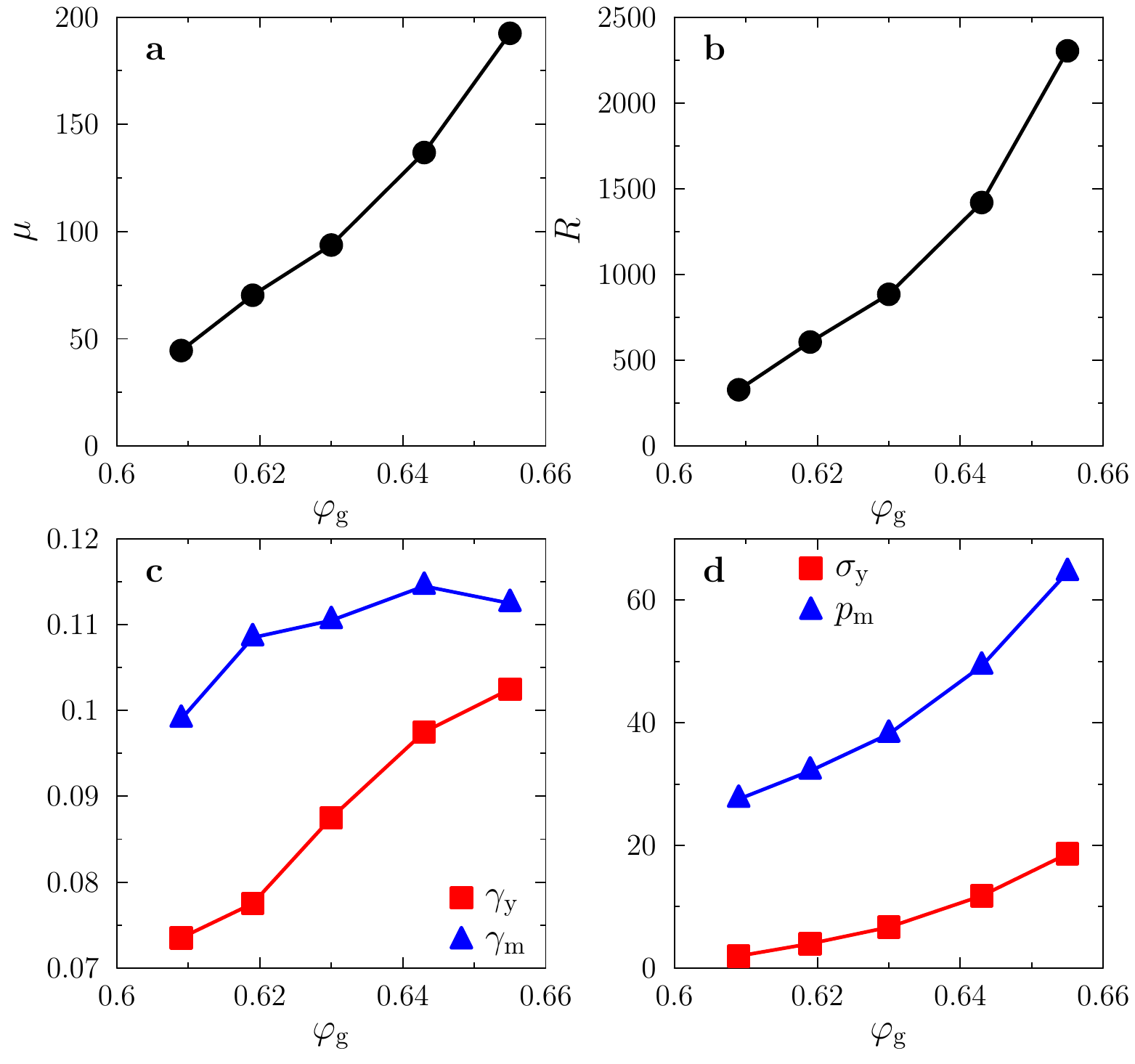}}}
\caption{Elastic and yielding parameters of equilibrium configurations. We plot as functions of $\varphi_{\rm g}$,  (a) the shear modulus $\mu$, (b) the dilatancy parameter $R$, (c) the yield strain $\gamma_{\rm y}$ and the strain $\gamma_{\rm m}$ at the maximum pressure in the pressure-strain curve, and 
(d) the yield stress $\sigma_{\rm y}$ and the maximum pressure $p_{\rm m}$. 
%The yielding stress $\sigma_{\rm y}$ vanishes around the MCT crossover $\varphi_{\rm d}= 0.594(1)$. (b) The shear modulus $\mu$ and (c) the dilatancy $R$ as functions of $\varphi$.
}
\label{fig:parameters}
\end{figure}

\label{sec:shear2}
\begin{figure}[h]
\centerline{\hbox{\includegraphics[width=0.6\columnwidth]{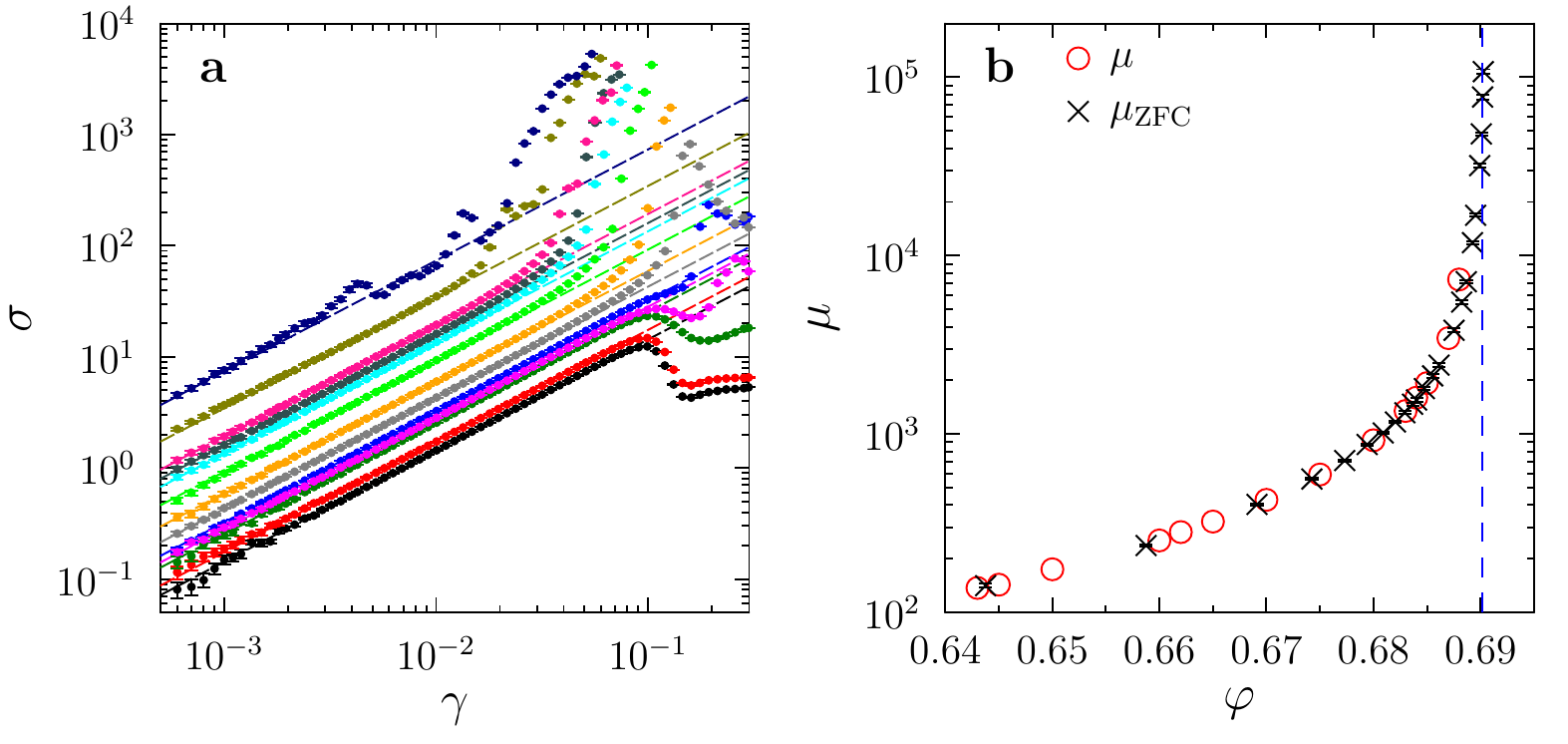}}}
\caption{Quasi-static shear on out-of-equilibrium glass states. The states are compressed from $\varphi_{\rm g} = 0.643$ to target density $\varphi$ before the shear is applied. The systems consist of $N=1000$ particles, and the data are averaged over $N_{\rm s} = 100$ samples and $N_{\rm th} = 10-30$ realizations for each sample. (a) The shear stress $\sigma$ as a function of $\gamma$, for a few different $\varphi$ (from bottom to top, $\varphi = 0.645, 0.65, 0.66, 0.662, 0.665, 0.67, 0.675, 0.68, 0.683, 0.684, 0.685, 0.687, 0.688$), \modi{where $\varphi_{\rm G}= 0.684(1)$}. The linear response regime is fitted to $\sigma = \mu \gamma$ (lines). (b) The shear modulus $\mu$ obtained from this fitting is compared to the ZFC shear modulus $\mu_{\rm ZFC}$ (we use $\gamma = 2 \times 10^{-3}$, see the main text), both of which diverge approaching to the jamming limit $\varphi_{\rm J} = 0.690(1)$ (vertical dashed line). For all figures, the error bars denote the standard error of the mean.}
\label{fig:comp_shear}
\end{figure}

\begin{figure}[h]
\centerline{\hbox{\includegraphics[width=0.3\columnwidth]{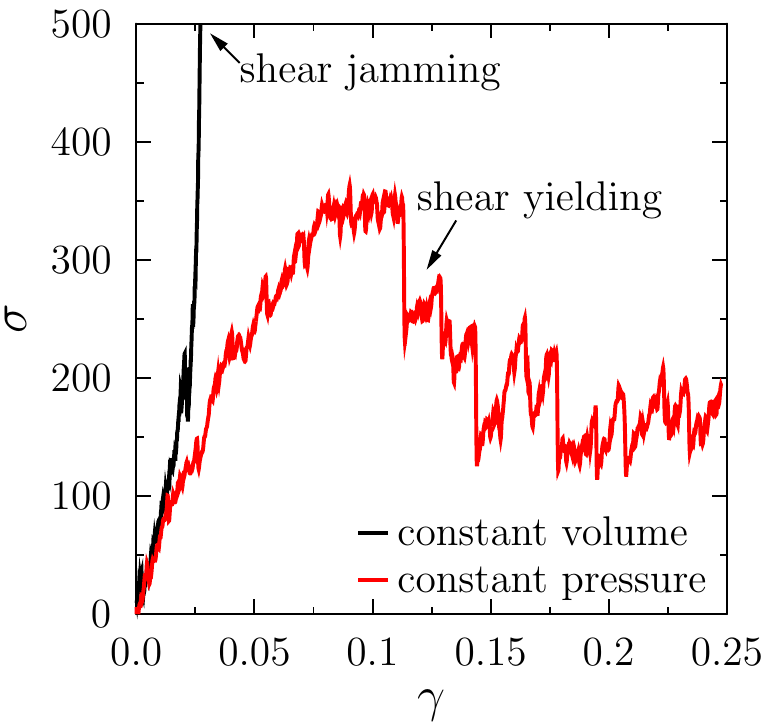}}}
\caption{Quasi-static shear on the same glass state ($\varphi_{\rm g}=0.643$, $\varphi = 0.688$, and $N=1000$) under constant volume and constant pressure show different behaviors at large $\gamma$. We observe shear jamming in the constant volume simulation and shear yielding in the constant pressure simulation. }
\label{fig:constP}
\end{figure}

\begin{figure}[h]
\centerline{\hbox{\includegraphics[width=0.6\columnwidth]{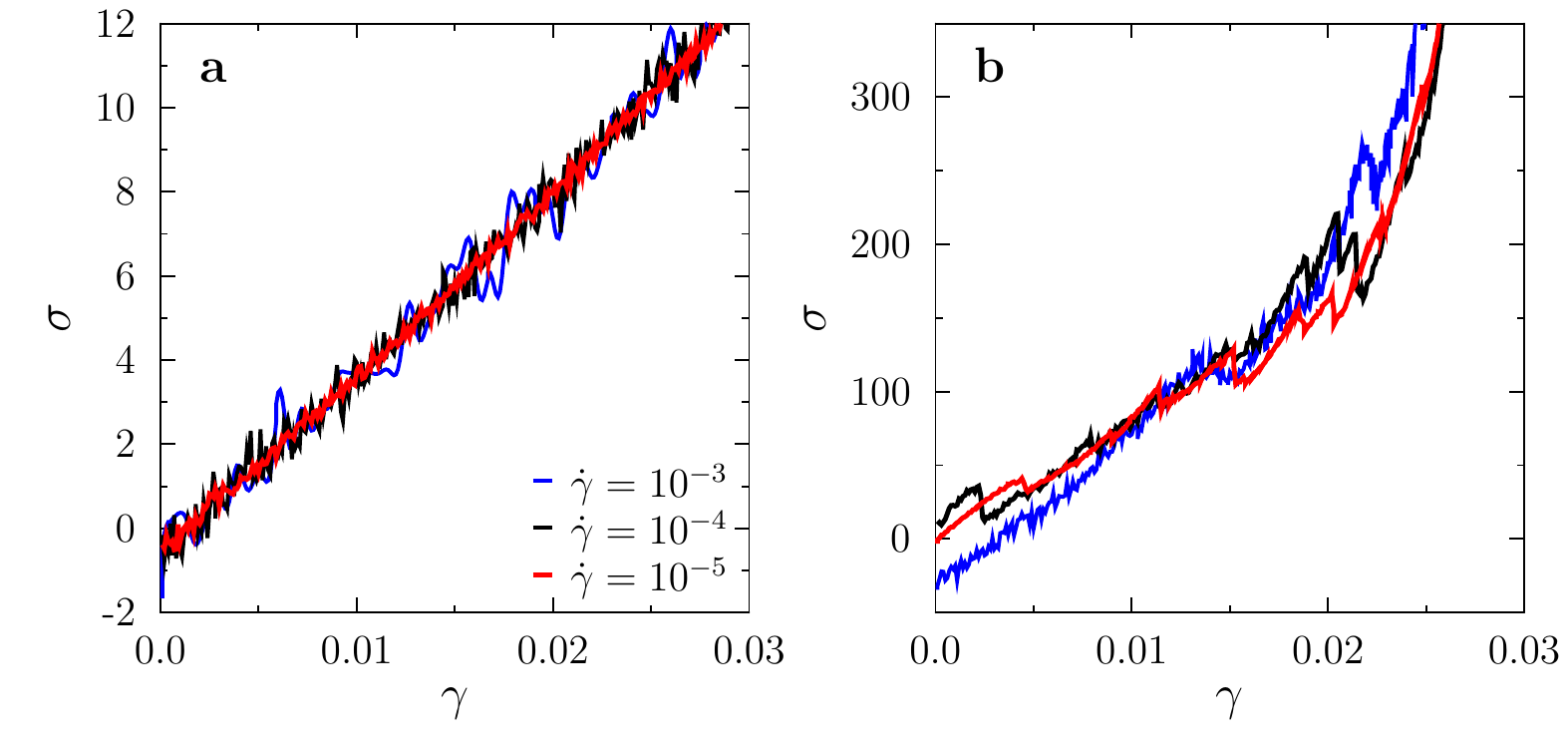}}}
\caption{
   The stress-strain curves of three {\it different realizations} of the compressed glass
at (a) $\varphi = 0.670$, and (b) $\varphi = 0.688$,
obtained from the {\it same equilibrated sample} of $N=1000$ particles at $\varphi_{\rm g}=0.643$.
They are driven by {\it different strain rates} $\dot{\gamma}$ as indicated by the legend.
}
\label{fig:shear_rate_dependence}
\end{figure}

\begin{figure}[h]
\centerline{\hbox{\includegraphics[width=0.6\columnwidth]{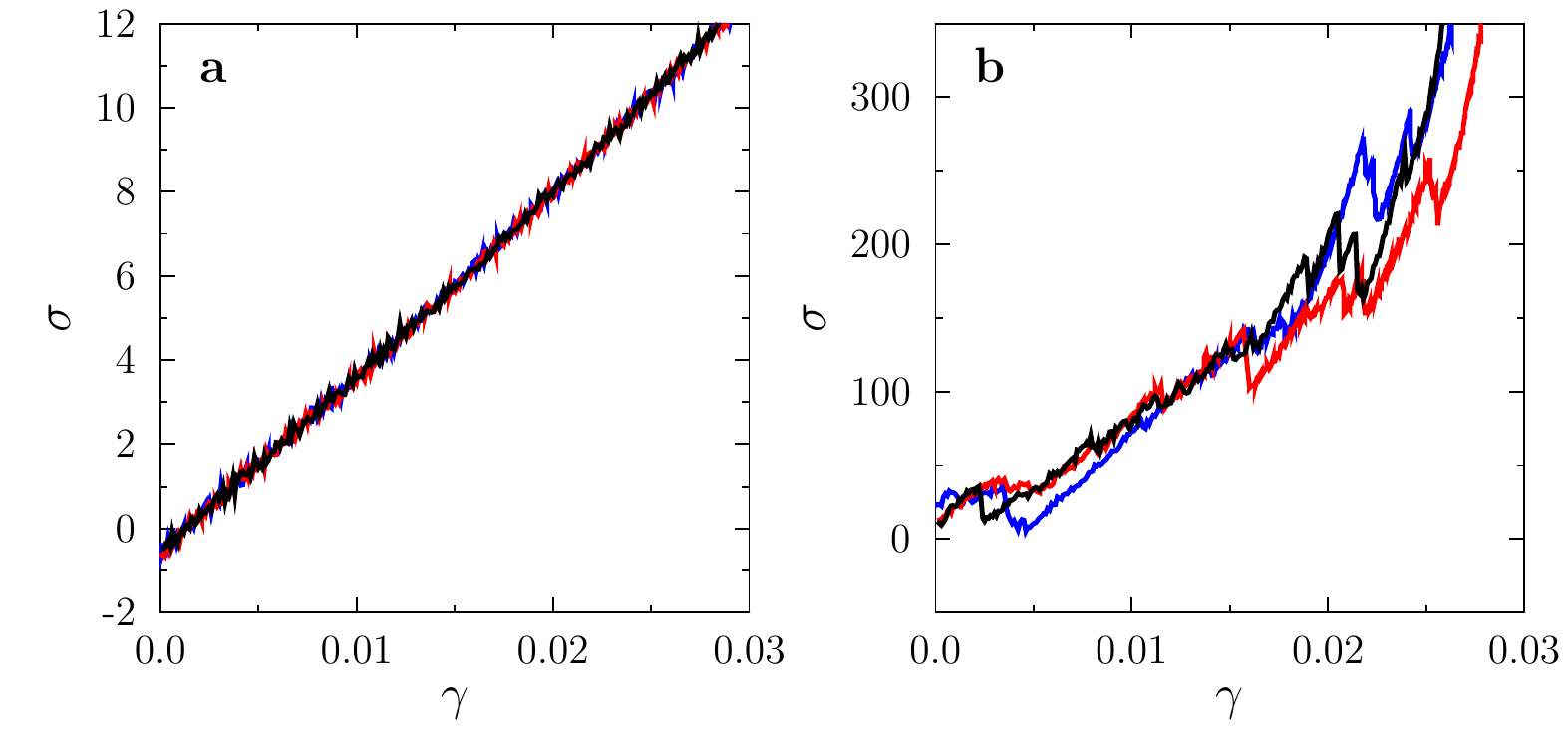}}}
\caption{
  Stress-strain curves of three {\it different realizations} of the compressed glass
    at (a) $\varphi = 0.670$, and (b) $\varphi = 0.688$,
obtained from the {\it same equilibrated sample} of $N=1000$ particles at $\varphi_{\rm g}=0.643$.
They are driven by the 
%{\it same strain rate}
\modi{\it common strain rates}
$\dot{\gamma} = 5 \times 10^{-6}$ for $\varphi =0.670$, and $\dot{\gamma} = 10^{-4}$ for $\varphi = 0.688$.}
\label{fig:realization_dependence}
\end{figure}

\begin{figure}[h]
\centerline{\hbox{\includegraphics[width=0.6\columnwidth]{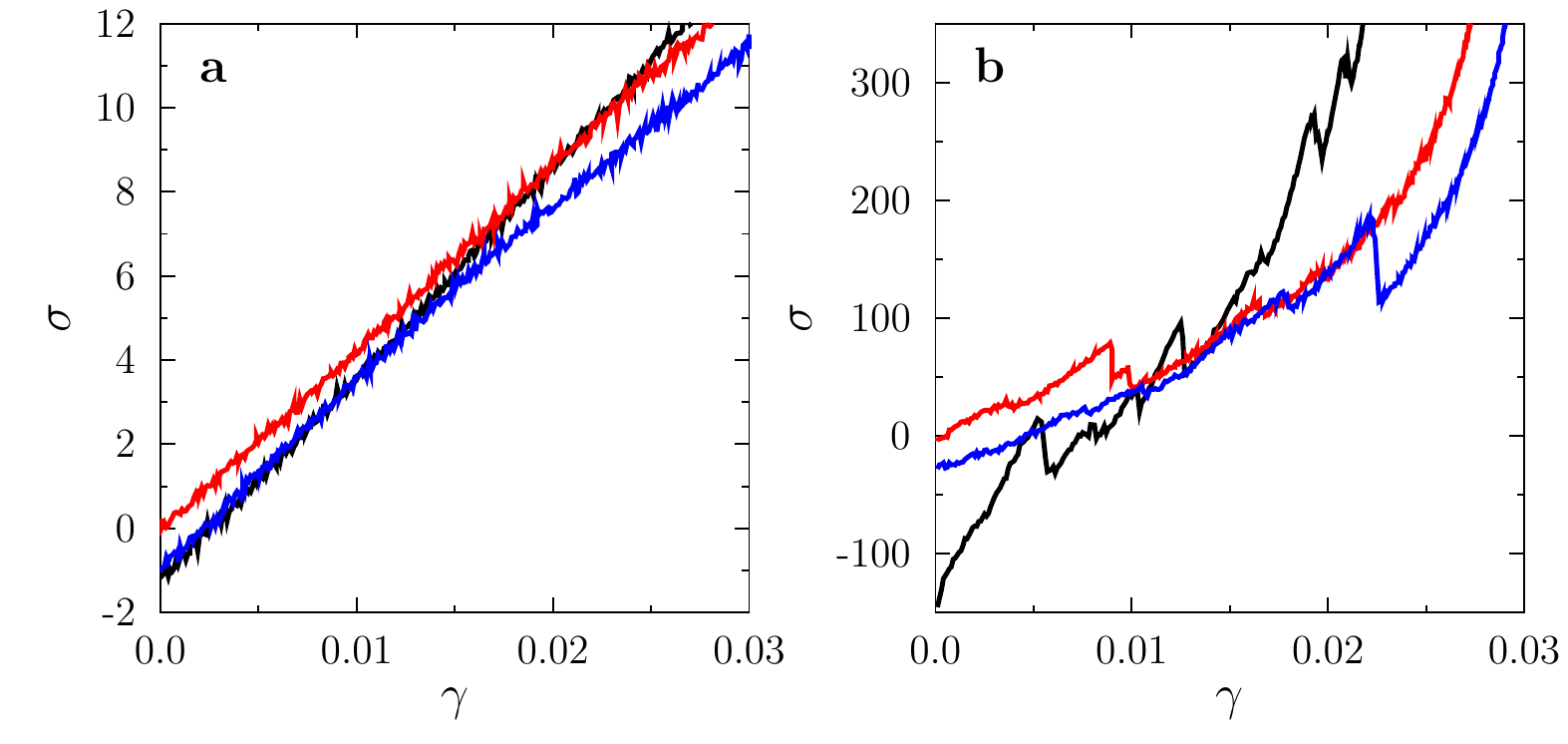}}}
\caption{Stress-strain curves on the compressed glasses at (a) $\varphi = 0.670$, and (b) $\varphi = 0.688$,
    obtained from three {\it different equilibrated samples} of $N=1000$ particles at $\varphi_{\rm g}=0.643$.
  They are driven by the 
  %{\it same strain rate}
  \modi{\it common strain rates}
$\dot{\gamma} = 5 \times 10^{-6}$ for $\varphi =0.670$, and $\dot{\gamma} = 10^{-4}$ for $\varphi = 0.688$.
}
\label{fig:sample_dependence}
\end{figure}

\begin{figure}[h]
\centerline{\hbox{\includegraphics[width=0.4\columnwidth]{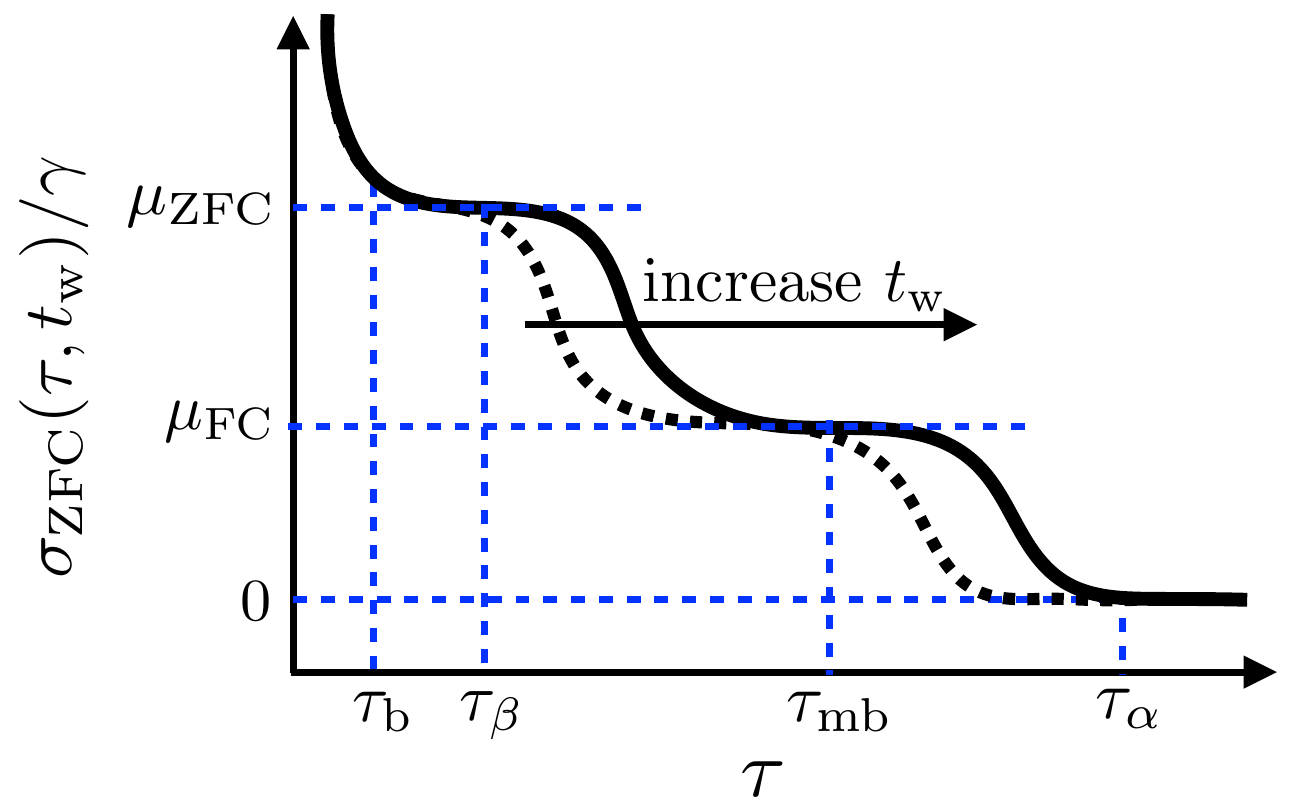}}}
\caption{Schematic illustration of the relaxation of ZFC shear stress $\sigma_{\rm ZFC}(\tau, t_{\rm w})$ after an instantaneous shear strain $\gamma$ is applied, in the Gardner phase $\varphi > \varphi_{\rm G}$~\cite{YZ2014PRE}. The two shear moduli $\mu_{\rm ZFC}$ and $\mu_{\rm FC}$ correspond to the first and second plateaus respectively. Corresponding time scales for the black solid line are indicated. The dotted black line represents a shorter waiting  time $t_{\rm w}$.
\modi{See also Fig. 2 of ~\cite{YZ2014PRE}.}
}
\label{fig:dynamics_illustration}
\end{figure}

\begin{figure}
\centerline{\hbox{\includegraphics[width=0.6\columnwidth]{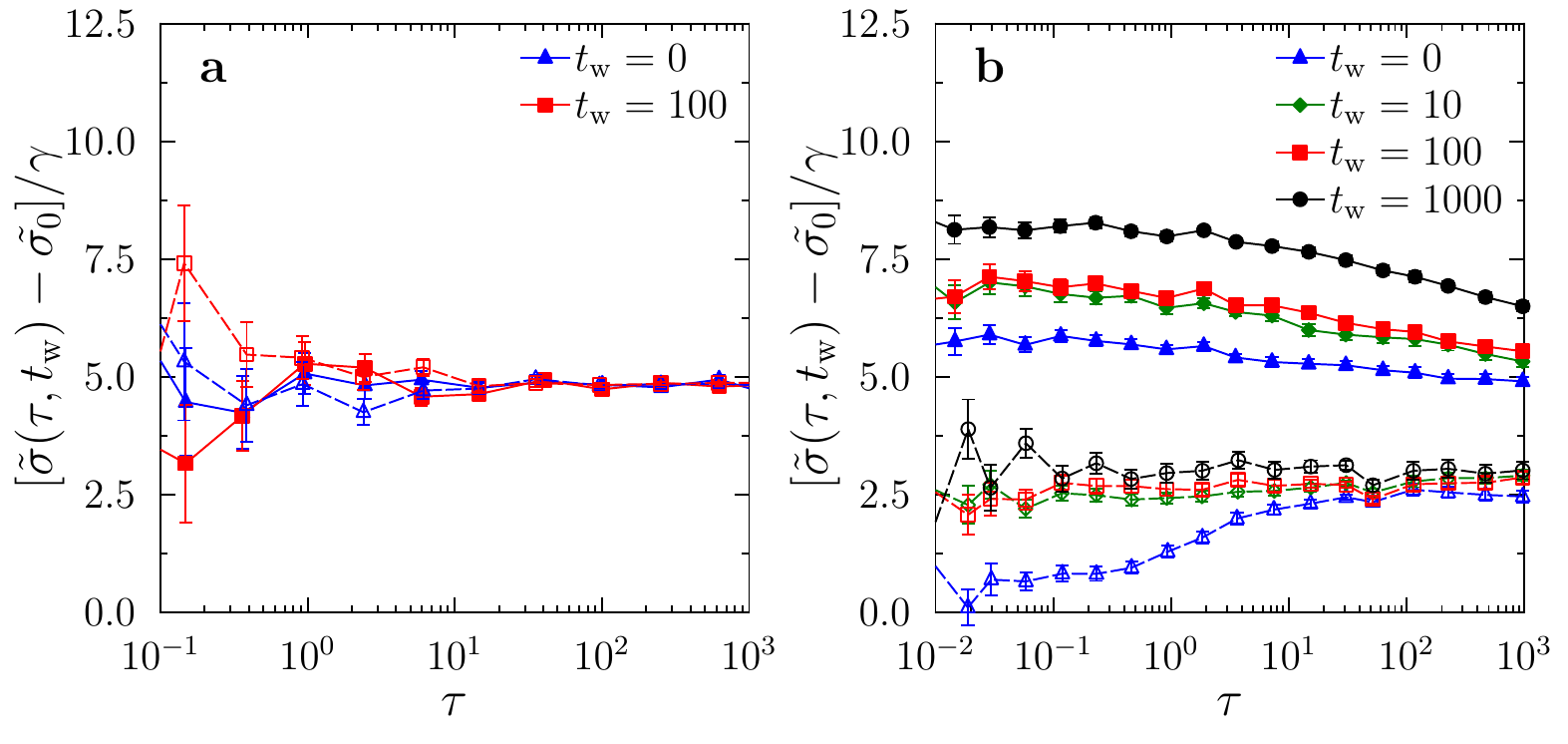}}}
\caption{Relaxation of $\tilde{\sigma}_{\rm ZFC} = \sigma_{\rm ZFC} /p $ (filled symbols)  and  $\tilde{\sigma}_{\rm FC} = \sigma_{\rm FC}/p$ (open symbols) at (a) $\varphi = 0.670$ and (b) $\varphi = 0.688$, for a few different $t_{\rm w}$, under a quasi-static shear strain $\gamma = 10^{-3}$. The system consists of $N=1000$ particles, and is compressed from $\varphi_{\rm g}  =0.643$. The  data are obtained for one individual sample, but averaged over  $N_{\rm th} \sim 1000$ independent realizations
  of the compressed glass.
  The remanent stress $\tilde{\sigma}_0$ has been subtracted from $\tilde{\sigma}$. } 
\label{fig:relax}
\end{figure}

\begin{figure}[h]
\centerline{\hbox{\includegraphics[width=0.6\columnwidth]{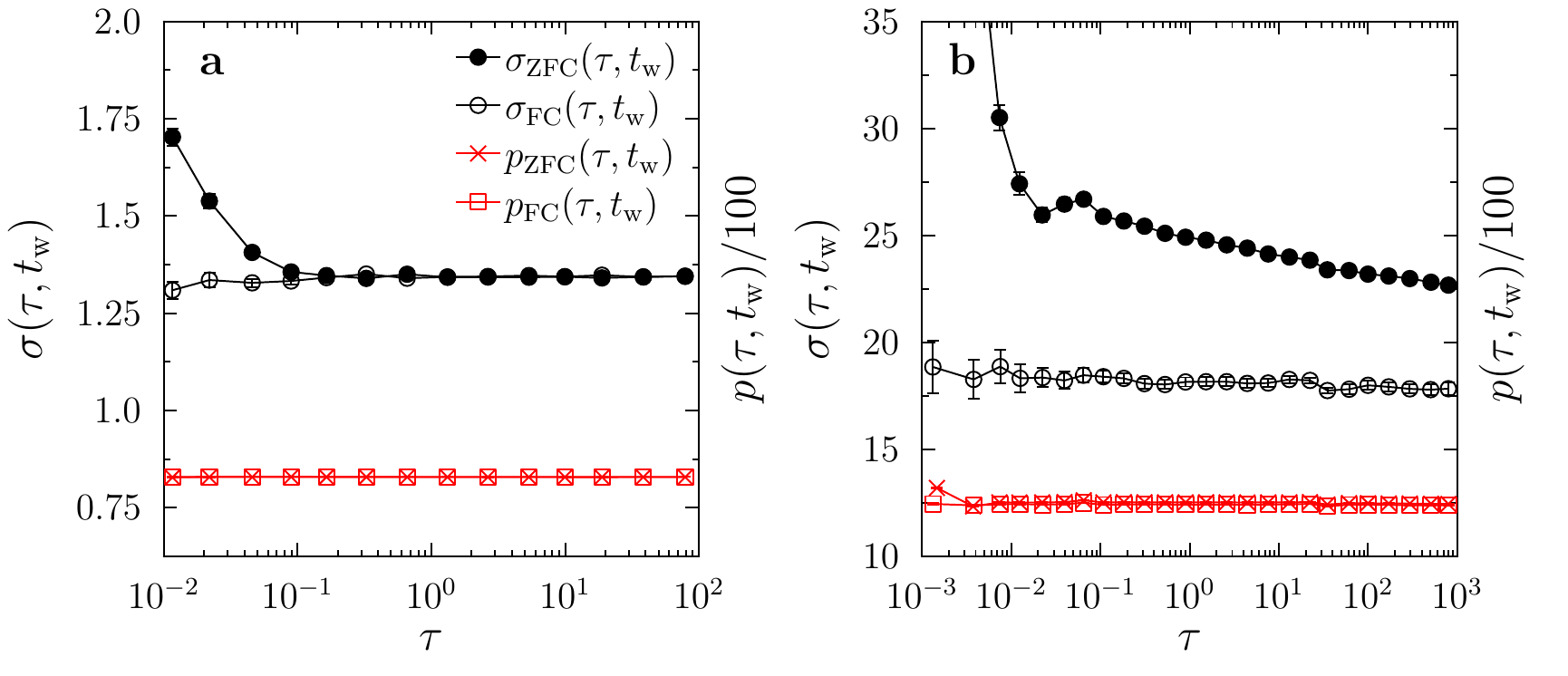}}}
\caption{\modi{Time evolution of the ZFC pressure $p_{\rm ZFC}(\tau, t_{\rm w})$ and the FC pressure $p_{\rm FC}(\tau, t_{\rm w})$  after an instantaneous increment of shear strain $\gamma = 10^{-3}$, at (a) $\varphi = 0.670$ and (b) $\varphi = 0.688$. \blue{For comparison, we also display the behaviour of the
shear stress in the ZFC and FC protocols.}
      Data are averaged over many realizations of compressed glasses obtained from a single equilibrated sample at $\varphi_{\rm g} = 0.643$ with $N = 1000$ particles. The pressure values are rescaled by a factor of $1/100$. This plot shall be compared with Fig. 2 in the main text (only $t_{\rm w} = 1000$  data are shown).}}
\label{fig:pressure}
\end{figure}

\begin{figure}[h]
\centerline{\hbox{\includegraphics[width=0.6\columnwidth]{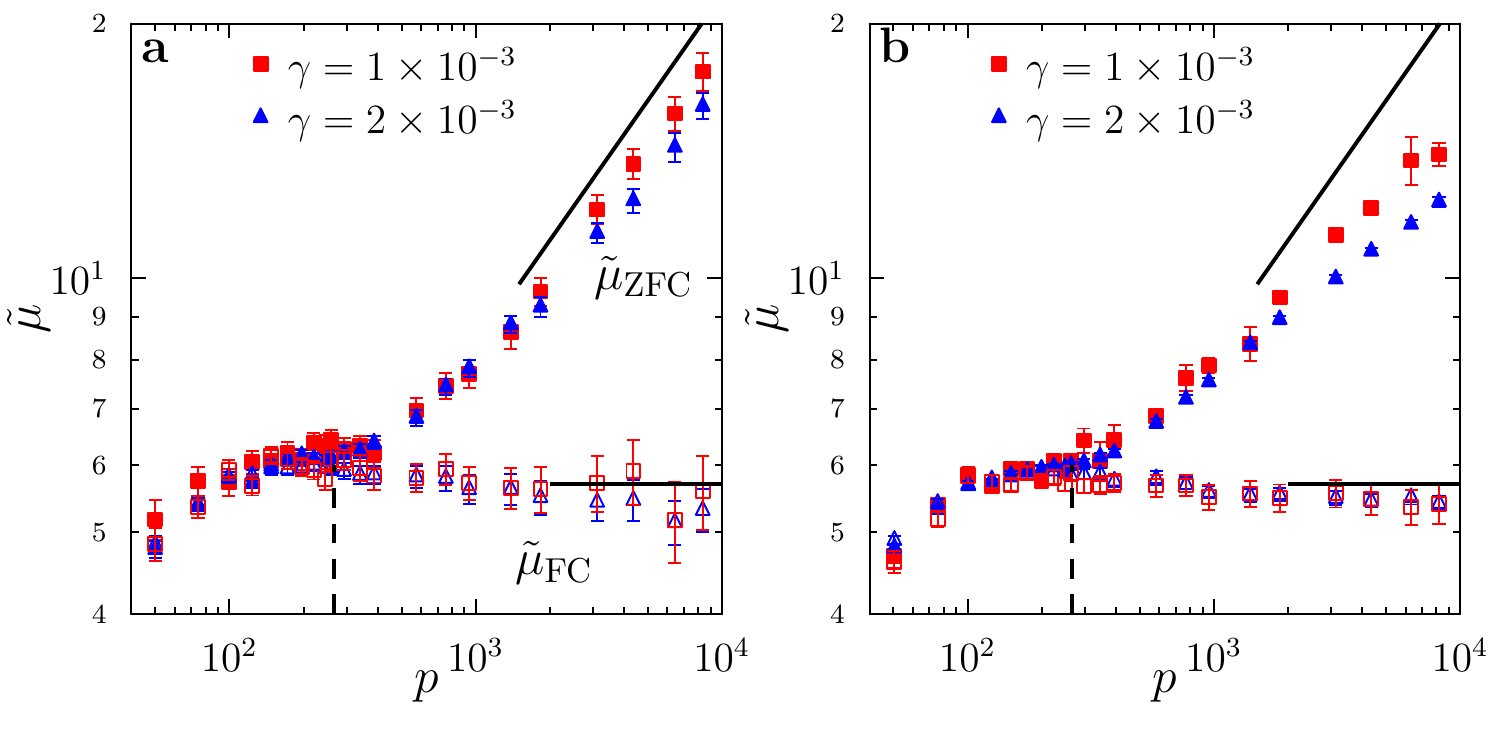}}}
\caption{$\gamma$-dependence on the ZFC and FC shear moduli. The data  are obtained for (a) $N=500$ and  (b) $N=2000$ particles, and are averaged over $N_{\rm s} \approx 200$ samples and $N_{\rm r} \approx 100$ 
individual realizations for each sample. The vertical dashed line represents the Gardner transition~\cite{BCJPSZ2016PNAS}, and the solid lines are the mean-field predictions $\mu_{\rm ZFC} \sim p^{1.41574}$ and $\mu_{\rm FC} \sim p$~\cite{YZ2014PRE}. } 
\label{fig:gamma_dependence}
\end{figure}

\begin{figure}[h]
\centerline{\hbox{\includegraphics[width=0.6\columnwidth]{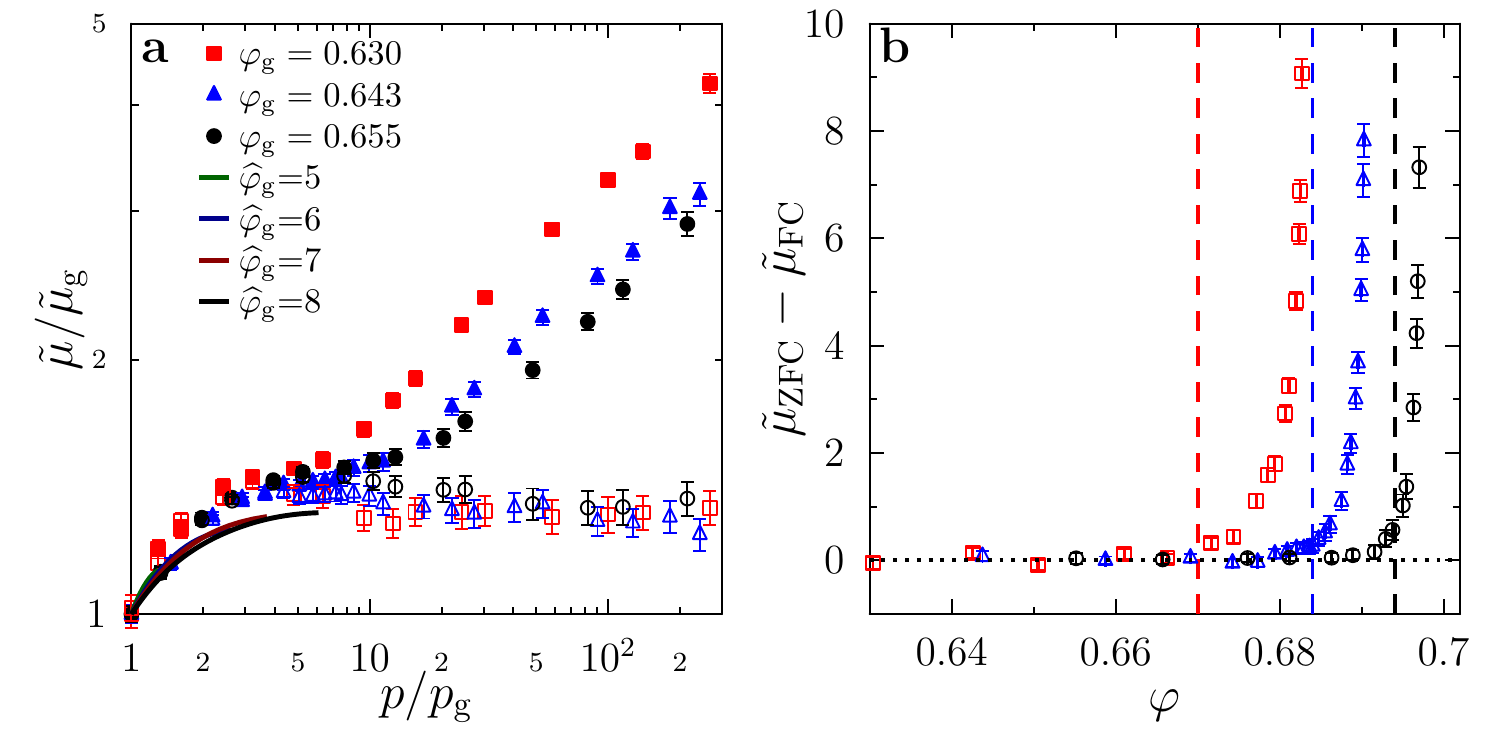}}}
\caption{$\varphi_{\rm g}$-dependence on the ZFC and FC shear moduli. The data  are obtained for $N=1000$ particles, and are averaged over $N_{\rm s} \approx 200$ samples and $N_{\rm r} \approx 100$ 
individual realizations for each sample. To compute the shear modulus, $\gamma = 2 \times 10^{-3}$ is used.
(a) The numerical data of the rescaled ZFC shear modulus $\tilde{\mu}_{\rm ZFC} = \mu_{\rm ZFC} /p$ (filled symbols) and the rescaled FC shear modulus $\tilde{\mu}_{\rm FC} = \mu_{\rm FC} /p$ (open symbols)  with a few different $\varphi_{\rm g}$,  are compared to the mean-field theory (lines) with different $\hat{\varphi}_{\rm g} $, where  $\hat{\varphi}_{\rm g} = 2^d \varphi_{\rm g} /d$ \blue{ with $d=3$}, following the convention used in Ref.~\cite{RUYZ2015PRL}. Both numerical and theoretical results are rescaled by the reference values $\tilde{\mu}_{\rm g} = \mu_{\rm g}/p_{\rm g}$ and $p_{\rm g}$ at $\varphi_{\rm g}$.  (b) The difference $\tilde{\mu}_{\rm ZFC} - \tilde{\mu}_{\rm FC}$ as a function of $\varphi$ for a few different $\varphi_{\rm g}$, where the Gardner transitions $\varphi_{\rm G}$ (values from Ref.~\cite{BCJPSZ2016PNAS}) are marked by vertical lines.} 
\label{fig:phig_dependence}
\end{figure}

\begin{figure}[h]
\centerline{\hbox{\includegraphics[width=0.3\columnwidth]{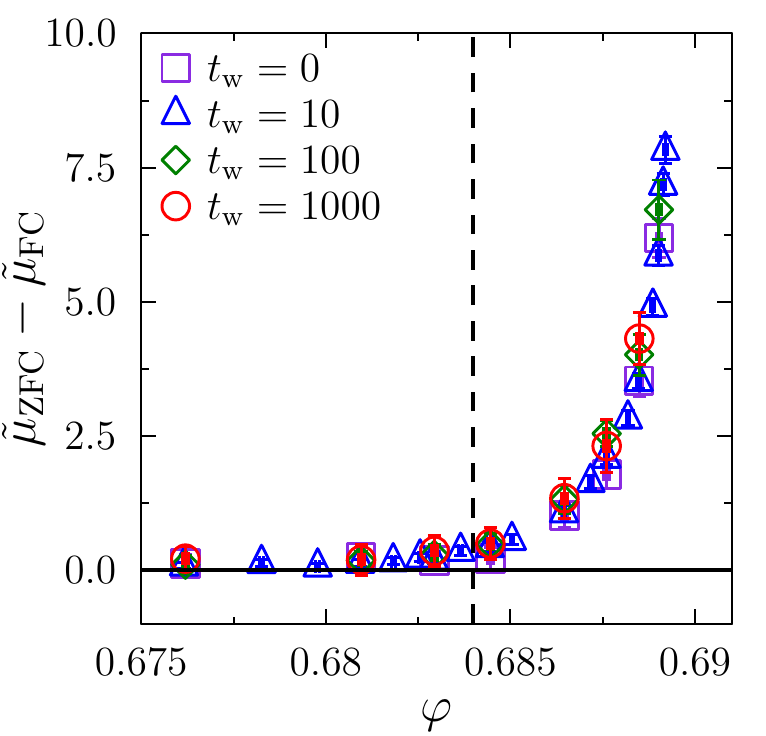}}}
\caption{\modi{The difference $\tilde{\mu}_{\rm ZFC} - \tilde{\mu}_{\rm FC}$ is plotted as a function of $\varphi$, for $\varphi_{\rm g} = 0.643$ and $N=1000$. The shear moduli are measured at  $\tau = 1$ for a few different waiting time $t_{\rm w}$. The data are obtained by using $\gamma = 2 \times 10^{-3}$, and are averaged over $N_{\rm s} \approx 100$ samples and $N_{\rm r} \approx 50$ independent realizations for each sample. The vertical dashed line represents $\varphi_{\rm G} = 0.684$~\cite{BCJPSZ2016PNAS}.} }
\label{fig:tw_dependence}
\end{figure}
\begin{figure}
\centerline{\hbox{\includegraphics[width=0.3\columnwidth]{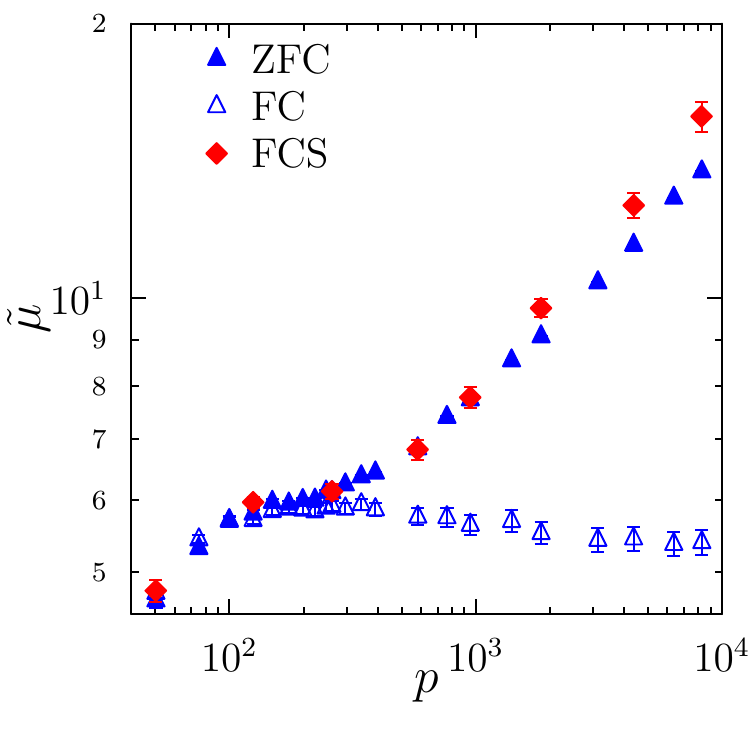}}}
\caption{The reduced shear modulus obtained from the third protocol (FCS) is compared to $\tilde{\mu}_{\rm ZFC}$ and $\tilde{\mu}_{\rm FC}$, for  $\varphi_{\rm g} = 0.643$, $N=1000$, and $\gamma = 2 \times 10^{-3}$.  The data are averaged over $N_{\rm s} \approx 200$ samples and $N_{\rm r} \approx 100$ individual realizations for each sample. } 
\label{fig:FCS}
\end{figure}

\begin{figure}[h]
\centerline{\hbox{\includegraphics[width=0.3\columnwidth]{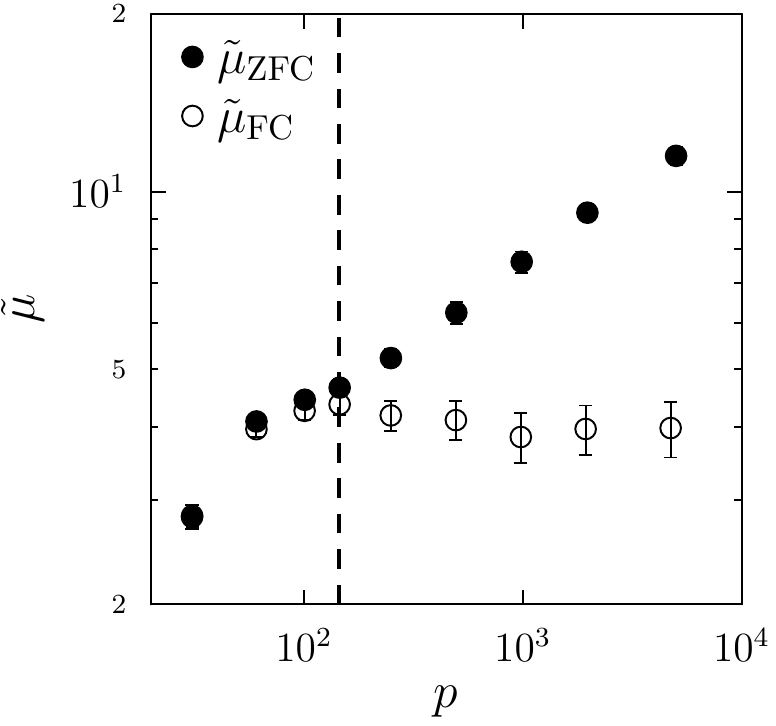}}}
\caption{Protocol-dependent shear modulus of a bidisperse hard disk glass former, where the vertical dashed line marks the Gardner transition estimated independently in Ref.~\cite{BCJPSZ2016PNAS}. The data are obtained for $\varphi_{\rm g} = 0.808$, and are averaged over $N_{\rm  s} \approx 100$ samples and $N_{\rm th} \approx 100$ realizations for each sample. }
\label{fig:2D}
\end{figure}

\clearpage

\section*{Supplementary Note 1 -- Model and observables}

\subsection*{\blue{Polydisperse hard sphere system}}

We study an assembly of $N$ polydisperse \modi{HSs}
whose diameters are distributed according to a probability distribution~\cite{BCNO2016PRL,BCJPSZ2016PNAS},
\beq
P(D) \sim D^{-3}, \qquad D_{\rm min} \leq D < D_{\rm min}/0.45.
\label{eq:PD}
\eeq
\modi{This distribution is chosen to optimize the swap algorithm so that denser equilibrium configurations can be obtained, while ensuring that crystallization is suppressed~\cite{BCNO2016PRL}.}
%We define the mean particle
%diameter as  $\bar{D}=\int_{D_{\rm min}}^{D_{\rm min}/0.45}dD P(D) D$
%and use it as the unit for the length scales.
The control parameter is the number density
$\rho = N/V$ or the volume fraction 
$\varphi=(\pi/6)\rho \int_{D_{\rm min}}^{D_{\rm min}/0.45}dD P(D) D^3$,
%$\varphi=(4\pi/3)\rho \bar{D}^{3}$.
%\blue{[ Would you put the exact expression?]}
where $V$ is the volume of the system.
The mode-coupling theory (MCT) dynamical crossover density is $\varphi_{\rm d} = 0.594(1)$~\cite{BCJPSZ2016PNAS}. 
The simulation time $t$ is expressed in units of $\sqrt{\beta m \bar{D}^2}$, where the inverse temperature $\beta$, the particle mass $m$, and the mean particle diameter   $\bar{D}=\int_{D_{\rm min}}^{D_{\rm min}/0.45}dD P(D) D$ all  set to unity.

\subsection*{\blue{Shear stress and pressure}}

For a HS system, the stress is entropic. The stress tensor  is given by
\beq
\Sigma_{mn} = -\frac{1}{V} \sum_{i<j} \vec{r}_{ij,m} \vec{f}_{ij,n}
\eeq
where $\vec{r}_{ij,m}$ is the $m$-th component of the separation vector $\vec{r}_{ij} = \vec{r}_i - \vec{r}_j$ between particles $i$ and $j$, and $\vec{f}_{ij, n}$ is the $n$-th component of the  inter-particle force  $\vec{f}_{ij}$. The force $\vec{f}_{ij}$ is computed from the exchange rate of the momentum between $i$ and $j$.
% $\vec{f}_{ij} = \frac{dp_{ij}}{dt}$. 
In our shear protocols, we are interested in the $z$-$x$ element of the stress tensor (we omit the subscript),
\beq
\Sigma =- \frac{1}{V} \sum_{i<j} z_{ij} \vec{f}_{ij,x}.
\eeq
The pressure $P$ is the negative average of three diagonal elements of the stress tensor, i.e., $P = -(\Sigma_{xx} + \Sigma_{yy} + \Sigma_{zz})/3 = \frac{1}{3V} \sum_{i<j} \vec{r}_{ij} \cdot \vec{f}_{ij}$. In this study, we report results in the units of reduced pressure 
$p = \beta P/\rho$
and reduced stress
$
\sigma = \beta \Sigma /\rho$.

\section*{Supplementary Note 2 -- Quasi-static shear on equilibrium configurations}
First let us report data obtained by quasi-static shear on {\it equilibrium} configurations at a few different $\varphi_{\rm g}$ (Supplementary Figure~\ref{fig:SF_phig}).
As we noted in the main text, the system  is in the liquid state in the thermodynamic sense (the Kauzmann density $\varphi_{\rm K}$, if any, is larger then $\varphi_{\rm g}$), but the $\alpha$-relaxation time is much larger than our simulation time scales so that the system behaves as a solid. The stress-strain curve, \modi{averaged over many samples and realizations}, shows a linear elastic regime at small $\gamma$, followed by yielding. We define the location of the peak in the stress-strain curve as the yield strain $\gamma_{\rm y}$. \modi{Note that the definition of $\gamma_{\rm y}$ is more ambiguous for the stress-strain curve of a single realization from a single sample (for example, see Fig.~1b in the main text). In this study, we do not attempt to precisely determine $\gamma_{\rm y}$ for each single stress-strain curve.} The shear modulus $\mu$ is determined by $\mu = \sigma/\gamma$ in the elastic regime. Both $\gamma_{\rm y}$ and $\mu$ grow with $\varphi_{\rm g}$  (Supplementary Figure~\ref{fig:parameters}).
The yield strain $\gamma_{\rm y}$ and the yield stress $\sigma_{\rm y}$ \modi{(which is the stress at 
  $\gamma_{\rm y}$)} appear to vanish continuously around $\varphi_{\rm g} \approx \varphi_{\rm d}$.
  On the other hand, the shear modulus $\mu$ appears to remain finite at $\varphi_{\rm d}$, which implies \modi{a} discontinuous jump of $\mu$ at $\varphi_{\rm d}$ being consistent  with the mean-field theory \cite{YZ2014PRE}.
  In the elastic regime, the dilatancy effect is observed: the pressure $p$ increases quadratically with $\gamma$,  i.e., $p=p_{\rm g} + R \gamma^2$, where $p_{\rm g}$ is the pressure at $\varphi_{\rm g}$ and $\gamma = 0$, and $R$ is the dilatancy parameter. The onset of the peak in the pressure-strain curves lags behind the yielding\modi{, i.~e. the peak of the stress-strain curves}. We compare our numerical data to the mean-field theoretical prediction~\cite{RUYZ2015PRL},  and find reasonable agreement on rescaled plots as shown in Supplementary Figure~\ref{fig:SF_phig}.

\section*{Supplementary Note 3 -- Quasi-static shear on out-of-equilibrium configurations}

\subsection*{Stress-strain curves}

Next, let us present quasi-static shear data of {\it out-of-equilibrium} configurations. These configurations are obtained by compressing the equilibrium configurations from $\varphi_{\rm g}$ to a target density $\varphi$, at a constant compression rate $\delta_{\rm g} = 10^{-3}$. Supplementary Figure~\ref{fig:comp_shear}a shows that at small $\gamma$, the average stress-strain curve has a linear regime, which shrinks  with increasing $\varphi$. Note that the data presented here are obtained by averaging over many samples and realizations, while the data in Fig.~1 (main text)  are for one single sample and one single realization.  For $\varphi > \varphi_{\rm G}$, the apparent linear regime in Supplementary Figure~\ref{fig:comp_shear}a is not truly elastic, since it is averaged over many  \modi{mesoscopic plastic events (MPEs)}  (Fig.~1). Thus the shear modulus defined as $\mu = \sigma/\gamma$ in the linear regime is not only due to purely harmonic responses but also involves non-affine corrections due to the plastic events. With this point being clarified, we show that the shear modulus $\mu$ obtained from fitting the data in the linear regime, is \modi{basically} consistent with $\mu_{\rm ZFC}$ presented in the main text (Supplementary Figure~\ref{fig:comp_shear}b).

At larger $\gamma$, we find that with increasing $\varphi$, the shear yielding disappears and the shear jamming emerges (Supplementary Figure~\ref{fig:comp_shear}a), which can be also observed in Fig.~1. Note that the simulation is performed under the constant volume condition. If we instead fix the pressure and allow the volume to change, then the shear jamming does not appear and  the shear yielding exists, even at large $\varphi$  (Supplementary Figure~\ref{fig:constP}). We also stress that the shift from the shear yielding to  the shear jamming is not correlated to the Gardner transition. In fact, Supplementary Figure~\ref{fig:constP} shows that it is possible to observe both \modi{MPEs} (at small $\gamma$) and yielding (at large $\gamma$) in the same stress-strain curve. The key difference is that, after \modi{a MPE}, the system remains in the same basin although it escapes from the sub-basin, and therefore it still  behaves like a solid, while after yielding, the system escapes from the basin and essentially behaves like a fluid. 
%In the former case, the shear stress $\sigma$ can still increase upon further increase of strain $\gamma$; in the latter case, it basically simply fluctuate around the constant average value. 

We next discuss in detail how the measurements of stress-strain curves depend on factors such as the compression rate $\delta_{\rm g}$, the shear rate $\dot{\gamma}$ (Supplementary Figure~\ref{fig:shear_rate_dependence}),  the realization (Supplementary Figure~\ref{fig:realization_dependence}), and  the sample (Supplementary Figure~\ref{fig:sample_dependence}). First of all, although these configurations are in principle out-of-equilibrium, they reach \modi{{\it restricted equilibrium}}~\cite{RUYZ2015PRL} for $\varphi < \varphi_{\rm G}$, i.e., they are nearly equilibrated within their glass basins. As shown in Ref.~\cite{BCJPSZ2016PNAS}, neither structural relaxation nor aging can be observed within the simulation time scale. According to that, in this regime, the results presented here should be nearly unchanged if a slower compression rate is used. The situation is different for $\varphi > \varphi_{\rm G}$: because the time scale diverges in this regime, it is difficult to obtain even the restricted equilibrium configurations and the data would be $\delta_{\rm g}$-dependent. Effectively, decreasing $\delta_{\rm g}$ is equivalent to increasing the waiting time $t_{\rm w}$. Since the $\delta_{\rm g}$-dependence has been well studied in previous work~\cite{BCJPSZ2016PNAS}, we do not repeat the analysis here. For other factors,  in the regime $\varphi < \varphi_{\rm G}$,  our results are independent of the shear rate (Supplementary Figure~\ref{fig:shear_rate_dependence}) and  realizations (Supplementary Figure~\ref{fig:realization_dependence}), although noticeable sample-to-sample variance (Supplementary Figure~\ref{fig:sample_dependence}) is observed.
%In contrast, for $\varphi > \varphi_{\rm G}$, each realization results in different stress-strain curves.}
In contrast, for $\varphi > \varphi_{\rm G}$, the stress-strain curve \modi{becomes} realization-dependent.
This observation is consistent with our basic expectation: the free energy landscape is complex in this regime, and the system could fall into different sub-basins after compression. 
%follow different paths on the free energy landscape under quasi-static shear. 

\subsection*{Relaxation of the shear stress: \blue{ connection to the  free-energy landscape}}
\label{sec:relax}

%\modi{
  %\sout{{\it Connection to the free energy landscape -- }}}
\modi{To better understand the relaxation of shear stress upon a instantaneous shear strain $\gamma$, in particular the behavior of  $\sigma_{\rm ZFC}(\tau, t_{\rm w})$ in the Gardner phase (Fig. 2b in the main text), for comparison we schematically plot $\sigma_{\rm ZFC}(\tau, t_{\rm w})$ anticipated  from a theoretical point of view~\cite{YZ2014PRE} (Supplementary Figure~\ref{fig:dynamics_illustration}). The key feature is that, after an initial fast decay within the ballistic time scale $\tau_{\rm b}$, two steps of relaxations  are expected:  the $\sigma_{\rm ZFC}(\tau, t_{\rm w})/\gamma$ firstly relaxes to the plateau corresponding to $\mu_{\rm ZFC}$ at time $\tau \sim \tau_\beta$, and then it further relaxes to the second plateau corresponding to $\mu_{\rm FC}$ at $\tau \sim \tau_{\rm mb}$. Here  $\tau_\beta$ and $\tau_{\rm mb}$ are the times needed to explore a single glass sub-basin and a glass meta-basin respectively. At even larger time $\tau \sim \tau_\alpha$, the stress may eventually relaxes to zero due to $\alpha$-processes. In our study, the initial configurations at $\varphi_{\rm g}$ are deeply equilibrated, such that $\tau_\alpha$ is far beyond our simulation time scale. Thus the $\alpha$-relaxation is irrelevant in our analysis. However, if the initial configurations are far away from equilibrium, then $\tau_\alpha$ could be comparable to the simulation time scale. In that case, the last step of relaxation towards zero stress may be observed~\cite{NYZ2015,OY13}. On the other hand, the behavior of $\sigma_{\rm FC}(t)$ is much simpler. As a one parameter function, it is $t_{\rm w}$-independent by definition, and $\sigma_{\rm FC}(t)/\gamma$ should converge quickly to $\mu_{\rm FC}$ after the initial ballistic processes.}

\subsection*{Relaxation of the shear stress: \blue{the case of slow switching on of the shear strain}}

%\modi{\sout{{\it The case of slow switching on of the shear strain -- }} }
In the main text we discussed the relaxation of the stress after instataneous shear.
\modi{Here let} us examine how the shear stress relaxes if a small shear strain $\gamma$ is applied {\it quasi-statically} (Supplementary Figure~\ref{fig:relax}). The data \modi{should be compared} with those in Fig.~2, where an instantaneous shear strain is applied. For $\varphi < \varphi_{\rm G}$, we do not see aging effects within our numerical accuracy.  The zero-field compression (ZFC) and the field compression (FC) shear stresses converge quickly to the same value. For  $\varphi > \varphi_{\rm G}$, no converge is observed within our simulation time window. The $\sigma_{\rm ZFC}(\tau, t_{\rm w})$ displays a plateau for short $\tau$, followed by slow decay.  Note that the time scale $\tau=1$ used in determining $\mu_{\rm ZFC}$ (see \modi{the} main text) is in the plateau region.

\subsection*{\blue{Time evolution of  the pressure}}
%\modi{\sout{{\it Time evolution of  the pressure -- }}
\modi{In the main text, we plot the rescaled shear stress $\tilde{\sigma} = \sigma/p$ in Figs. 2 and 3, since a simple scaling relation $\sigma \sim p$ is expected in the normal glass phase. Here we examine whether the pressure $p$ depends on time and protocol. Indeed, Supplementary Figure~\ref{fig:pressure} shows that, in contrast to the stress, the pressure is nearly time-independent and protocol-independent after instantaneous shear, both below and above the Gardner transition. Therefore,  $\tilde{\sigma}$ truly reveal the behavior of stress since we can treat $p$ as a constant at any $\varphi$.}

\section*{Supplementary Note 4 -- Additional data on the protocol-dependent shear modulus}
Here we report supplementary data on the protocol-dependent shear modulus. \modi{We discuss how the ZFC and the FC shear moduli depend on the the shear strain $\gamma$, the number of particles $N$, equilibrium density $\varphi_{\rm g}$,  and the waiting time $t_{\rm w}$.  
%We discuss the $\varphi_{\rm g}$-dependence on the ZFC and the FC shear moduli, 
%and provide the shear modulus obtained from 
We also measure the shear modulus using a third protocol.}

\subsection*{\modi{Dependence on the shear strain $\gamma$}}

\modi{
In our analysis, the shear modulus is measured by taking the ratio between the stress and the strain, i.e., $\mu_{\rm ZFC} =  (\sigma_{\rm ZFC} - \sigma_0)/\gamma$ and $\mu_{\rm FC} =  (\sigma_{\rm FC} - \sigma_0)/\gamma$, where $\sigma_0$ is the remanent shear stress. If $\gamma$ is sufficiently small such that the non-linear corrections are negligible,  the measured modulus should be independent of $\gamma$. Our data show that  the FC modulus $\mu_{\rm FC}$ is indeed in such a linear regime for the chosen $\gamma$ (see  Fig.~3b in the main text for $N=1000$ systems, and Supplementary Figure~\ref{fig:gamma_dependence} for $N=500$ and $N=2000$ systems). However, $\gamma$-dependence is observed for $\mu_{\rm ZFC}$: at large pressure $p$ close to jamming, $\mu_{\rm ZFC}$ increases for smaller shear strain $\gamma$. For smaller $\gamma$, the large-$p$ scaling $\mu_{\rm ZFC} \sim p^{\kappa}$ agrees better with the mean-field theory, for any $N$ studied, but additional data are required to conclude if the mean-field result is coincided in the limit $\gamma \to 0$. Recently, a very careful study shows that the mean-field jamming exponents, which characterize the critical distribution of small inter-particle gaps and weak contact forces, are consistent with simulation data in finite dimensions, after removing localized bucking excitations~\cite{CCPZ15}. Such an analysis in the  $p \to \infty$ limit is beyond the present numerical accuracy.
}

\modi{
\subsection*{\modi{Dependence on the number of particles $N$}}
For a fixed $\gamma$, Fig. 3 in the main text shows that at large $p$, $\mu_{\rm ZFC}$ decreases with increasing $N$. It suggests that the non-linear effect, associated to stress relaxation due to MPEs, is stronger in larger systems. Indeed, in Ref.~\cite{karmakar2010statistical}, the authors found a finite-size scaling
$\delta \gamma_1 \sim N^{\beta}$ with $\beta \approx -0.62$, for the mean strain $\delta \gamma_1$ at which the first MPE takes place in amorphous solids.
This scaling suggests that, MPEs are easier to occur in larger systems, and become unavoidable at any finite shear strain in the thermodynamic limit, because $\delta \gamma_1 \to 0$  as $N \to \infty$. It is thus reasonable to see that data with smaller
$\gamma$ (for a fixed $N$) or smaller $N$ (for a fixed finite $\gamma$)
obeys better the mean-field scaling $\mu_{\rm ZFC} \sim p^{1.41574}$, because the theoretical $\mu_{\rm ZFC}$
%\sout{exclude any plastic events for $\mu_{\rm ZFC}$} 
%\blue{
is only concerned about the linear response~\cite{YZ2014PRE}.}

\modi{In our data (Fig. 3b in the main text and Supplementary Figure~\ref{fig:gamma_dependence}), we do not find appreciable $N$-dependence of $\mu_{\rm FC}$. In contrast, a scaling relation $\mu_{\rm FC} \sim N^{-0.25}$ was reported in Ref.~\cite{NYZ2015}. Here we discuss possible reasons for this discrepancy. In~\cite{NYZ2015}, the systems are quenched from completely random initial configurations. Compared to the case in well equilibrated systems, in such non-equilibrium systems, the stress relaxes much faster and eventually decays  to zero, i.e., the system melts quickly~\cite{OY13} (see Supplementary Figure~\ref{fig:dynamics_illustration} for an illustration). Considering that larger systems have an easier tendency to relax, we expect that in the thermodynamic limit, the system turns to a liquid within the simulation time scale used in Ref.~\cite{NYZ2015}, which is the reason why $\mu_{\rm FC} \to 0$.
}

\subsection*{\modi{Dependence on the initial equilibrium density $\varphi_{\rm g}$}}
We find that our basic observation -- the bifurcation between the ZFC shear modulus $\mu_{\rm ZFC}$ and the FC shear modulus $\mu_{\rm FC}$ at the Gardner transition $\varphi_{\rm G}$ -- is independent of  the initial equilibrium density $\varphi_{\rm g}$ (see Supplementary Figure~\ref{fig:phig_dependence}). Note that the value of $\varphi_{\rm G}$ itself depends on  $\varphi_{\rm g}$. The large pressure ($\varphi \gg \varphi_{\rm G}$) scalings, 
$\mu_{\rm ZFC}$ versus $p$, and  $\mu_{\rm FC}$ versus $p$, 
are nearly unchanged for different $\varphi_{\rm g}$.
Additionally,  we compare our simulation data with theoretical predictions for $\varphi < \varphi_{\rm G}$.  We plot $\tilde{\mu}_{\rm ZFC}/\tilde{\mu}_{\rm g}$ and $\tilde{\mu}_{\rm FC}/\tilde{\mu}_{\rm g}$ as functions of $p/p_{\rm g}$ obtained from simulations, together with the mean-field {\it state following} theory~\cite{RUYZ2015PRL}, where  $\tilde{\mu}_{\rm} = \mu/p$ is the modulus rescaled by the pressure, and 
$\mu_{\rm g}$ and $p_{\rm g}$ are the shear modulus and the pressure at $\varphi_{\rm g}$. Note that the theory does not distinguish between ZFC and FC moduli in this regime.  On this rescaled plot, the theory and the simulation data show similar behaviors, both of which are insensitive to $\varphi_{\rm g}$. 
We point out  that the mean-field theory uses an over-simplified liquid EOS, that is only valid for mono-disperse hard spheres in the large dimensional limit. Thus a direct comparison between the theory and our simulation is impossible. 
However, once the effect of this mismatch on the liquid structure is removed by a proper rescaling with respect to the reference point at $\varphi_{\rm g}$,  the theory basically captures the general trend on how the system evolves under a slow compression annealing.
% (which resembles the state following procedure in the theory.

\modi{\subsection*{\modi{Dependence on the waiting time $t_{\rm w}$}}
In 
%Sect.~\ref{sec:relax}, 
\modi{Supplementary Note 3}, 
we have discussed how $\sigma_{\rm ZFC} (\tau, t_{\rm w})$ and  $\sigma_{\rm FC} (\tau, t_{\rm w})$ relax with $\tau$ under a quasi-static shear strain $\gamma$. Based on the data (Supplementary Figure~\ref{fig:relax}), we choose time scales $\tau = 1$  and $\t_{\rm w} = 10$ to measure $\mu_{\rm ZFC}$ and $\mu_{\rm FC}$ (see the main text). The scale $\tau = 1$ is chosen within the first plateau regime of $\sigma_{\rm ZFC} (\tau, t_{\rm w})$. Note that for larger $\tau$, the difference  $\mu_{\rm ZFC} -\mu_{\rm FC}$ would decrease, and would eventually vanish as $\tau \to \infty$, even in the Gardner phase. On the other hand, in order to examine the $t_{\rm w}$-dependence more carefully, we obtain additional data of $\mu_{\rm ZFC}$ and $\mu_{\rm FC}$ for a few different $t_{\rm w}$ (we fix $\tau =1$ in all cases). Interestingly, we found that the differences $\mu_{\rm ZFC} -\mu_{\rm FC}$ obtained by using  different $t_{\rm w}$, when plotted as a function of $\varphi$, collapse onto the same curve, within our numerical accuracy (Supplementary Figure~\ref{fig:tw_dependence}). In particular, because the Gardner transition is determined as the bifurcation point between $\mu_{\rm ZFC}$ and $\mu_{\rm FC}$, this result shows that the location of the transition point is independent of $t_{\rm w}$ in our approach. 
}

\subsection*{A third protocol}
To further verify and emphasize the protocol dependence on the shear modulus, we design a third protocol, in which we apply an additional shear strain after the FC procedure. In this protocol, we first apply a small quasi-static strain $\gamma$ at $\varphi_{\rm g}$,  compress the system to $\varphi$, and then after waiting for  $t_{\rm w} = 10$ and $\tau=1$, we measure the stress $\sigma_{\rm FC}$. 
This procedure is basically the same as the FC (see \modi{the} main text). We then apply an additional quasi-static strain $\gamma$  at $\varphi$, and measure the stress $\sigma_{\rm FCS}$ after waiting for $\tau=1$. The FCS (FC+shear) modulus is defined as $\mu_{\rm FCS} = (\sigma_{\rm FCS} - \sigma_{\rm FC})/\gamma $. Supplementary Figure~\ref{fig:FCS} shows that this shear modulus 
 is close to $\mu_{\rm ZFC}$, and clearly different from $\mu_{\rm FC}$. From the view point of free energy landscape, for $\varphi > \varphi_{\rm G}$,  $\mu_{\rm FCS}$ represents the local curvature of the sub-basins at a finite given $\gamma$ (see Fig. 4 \modi{in the main text}), while  $\mu_{\rm ZFC}$ is the local curvature of the sub-basins at $\gamma=0$.

\section*{Supplementary Note 5 -- Protocol-dependent shear modulus of bidisperse hard disks}
To test if the ZFC/FC approach can be applied to other systems, we also study a two-dimensional bidisperse hard disk model glass former. The system contains $N=1000$ equimolar bidisperse hard disks with diameter ratio $D_1:D_2 = 1.4:1$. The dynamical crossover density is $\varphi_{\rm d} = 0.790(1)$~\cite{BCJPSZ2016PNAS}. The example in Supplementary Figure~\ref{fig:2D} shows that the shear modulus becomes protocol-dependent for $\varphi > \varphi_{\rm G}$. This signature can be used to determine  $\varphi_{\rm G}$, whose value is fully consistent with the previous independent estimate~\cite{BCJPSZ2016PNAS}.

\section*{Supplementary Methods -- Numerical protocols}

%\subsection{Swap algorithm}
\subsection*{\modi{Compression protocols}}

\modi{{\it Generation of dense equilibrium liquids --} }
To prepare dense equilibrium configurations, we combine the Lubachevsky-Stillinger (LS) algorithm~\cite{SDST06} with swap Monte Carlo moves~\cite{BCNO2016PRL,BCJPSZ2016PNAS}.
\modi{The LS algorithm consists of standard event-driven molecular dynamics (MD), and slow compression
which is realized by variation of the diameter of particles.}
Our protocol consists of  the following two steps:
\begin{enumerate}
\item Starting from an ideal gas configuration, we first compress it  to $\varphi_0 = 0.54$, by growing spheres at a constant rate $\delta_{\rm g} = 10^{-3}$, such that $D(t) = D(0)(1+\delta_{\rm g} t)$.
%\blue{[Define time unit!]}
Because the process is equivalent to compression, hereafter we call $\delta_{\rm g}$  as compression rate.
% (rather than `growth rate').
%The density $\varphi_0 = 0.54$ is above the freezing point  of the liquid-crystal phase transition. 
This initial compression is fast enough to suppress crystallization, and slow enough to equilibrate the configuration up to $\varphi_0 $.

\item Starting from the equilibrium configuration at $\varphi_0$, we switch to a slower compression rate $\delta_{\rm g} = 10^{-5}$, and compress the configuration to a higher density $\varphi_{\rm g}$. Swap attempts are introduced:  we randomly pick a pair of particles and exchange their diameters if no overlap is created after the swap. The particle sizes do not change during swap moves. We perform $10\%$ swap moves and $90\%$ LS molecular dynamics steps. After the compression, we further relax the configuration for $t=1000$ to check that the pressure does not change. %The simulation time $t$ is expressed in units of $\sqrt{\beta m \bar{D}^2}$, where the inverse temperature $\beta$, the particle mass $m$, and the mean particle diameter $\bar{D}$ all all set to unity. 
The equation of state (EOS) of our equilibrium configurations agrees with  the Carnahan-Stirling (CS) expression (see Supplementary Figure~\ref{fig:swap})~\cite{B70},
\beq
p_{\rm CS}(\varphi) = \frac{1}{1-\varphi} + \frac{3 s_1 s_2}{s_3} 
\frac{\varphi}{(1-\varphi)^2} + \frac{s_2^3}{s_3^2} \frac{(3-\varphi)
\varphi^2}{(1-\varphi)^3},
\label{eq:poly_EOS}
\eeq
where $s_k$ is the $k$-th moments of the diameter distribution function $P(D)$.
%\modi{\sout{given by \eq{eq:PD}.}}

%\item To obtain equilibrium configurations at higher $\varphi_{\rm g}$, we compress a configuration equilibrated at  $\varphi_{\rm g}' < \varphi_{\rm g}$. This is more efficient than compressing a configuration directly from $\varphi = \varphi_0$. The growth rate $\delta_{\rm g}$ is adjusted to smaller values for higher $\varphi_{\rm g}$.

\end{enumerate}

\modi{{\it Generation of glasses --}}
The swap algorithm is swithced off once the
equilbriated configuration at the target $\varphi_{\rm g}$ is obtained.
  All the subsequent simulations are performed using the
  \modi{MD without swap}.
  Since the $\alpha$-relaxation time \modi{has become}  much larger
  than our MD simulation time scales, we are left with a piece of glass.
  \modi{We call the configuration of the particle positions $\{\vec{r}_{i}\}$ ($i=1,2,\ldots,N$)
    of such a glass at $\varphi_{\rm g}$ as a {\it sample}. From each of such a sample at $\varphi_{\rm g}$, we generate many
       {\it realizations} by setting random particle velocities $\{\vec{v}_{i}\}$ ($i=1,2,\ldots,N$)
       drawn from the Maxwell-Boltzmann disribution. Each of such realizations is compressed by
       the LS algorithm to obtain compressed glasses at desired densities $\varphi > \varphi_{\rm g}$.
  Note that the kinetic energy is conserved so that the system remains at the unit temperature throughout our simulations.}
  
\subsection*{Shear protocols}

{\it Quasi-static shear --}
In the quasi-static shear, the shear strain $\gamma$ is increased with time at a constant rate $\dot{\gamma}$, which is set small enough such that the system is quasi-equilibrated at each step. 
%An instantaneous and infinitesimal increase of $\gamma$ is followed by the conjugated gradient (CG) method to remove overlappings, and event-driven molecular dynamics (MD) simulations to equilibrate the system. The above two procedures are combined in such a way that the shear rate $\dot{\gamma}$ is a constant. More specifically, 
The protocol consists of the following steps:
\begin{enumerate}
\item Increase the shear strain $\gamma$ instantaneously by an infinitesimal amount $\gamma \to \gamma + \delta \gamma$ with $\delta \gamma = 10^{-4}$. We perform an affine deformation to all particles, whose positions are shifted by $x_i \to x_i + \delta \gamma z_i$, where $x_i$ and $z_i$ are the $x-$ and $z-$coordinates of particles $i$. 
%Here the particles are considered to be immersed in a (background) fluid field under shear. 
This instantaneous shift could introduce overlaps between some particles, which are removed by using the the conjugated gradient (CG) method~\cite{press2007numerical}. To use CG, a harmonic inter-particle potential $\phi_{ij}(r) =  (1-r/D_{ij})^2$ (zero when $r > D_{ij}$) is used, where $D_{ij}=(D_i + D_j)/2$ is the average diameter of particles $i$ and $j$. 
%We set $\epsilon$ equal to the mean diameter $\bar{\sigma}$, such that the inter-particle force is unitless. 
The boundary condition in the $z$ direction satisfies the Lees-Edwards scheme~\cite{lees1972computer}, i.e., the $x$-position of the top (bottom) imaginary box is shifted by $\delta \gamma L$ ($-\delta \gamma L$), where $L$ is the linear size of the simulation box. After this step, we obtain an non-overlapping  \modi{hard sphere (HS)} configuration under shear strain $\gamma + \delta \gamma$. %which are not necessarily equilibrated. 

\item We \modi{switch the soft potential back to the hardcore potential}, and equilibrate the system by using \modi{the event-driven MD}
  to simulate the dynamics of HSs under \modi{fixed}  shear strain $\gamma +  \delta \gamma$.
  \modi{Again we emphasize that the dynamics preserves the kinetic energy so that the system remains at a constant temperature.
  The velocities are reinitialized after each step of the shear strain.}
  The Lees-Edwards boundary condition is kept.
%, i.e., the top/bottom imaginary box is shifted by $+/-\gamma L$. The simulation time here is expressed in units of $\sqrt{\beta m \sigma^2}$, with unit inverse temperature $\beta$, unit particle mass $m$, and unit average particle diameter $\bar{\sigma}$. 
We perform LS simulation for a duration $\delta t$, such that $\delta \gamma / \delta t = \dot{\gamma}$.
\item The above two steps are repeated until the shear strain reaches a target value. 
%The shear rate $\dot{\gamma}$ is then set to zero, but the system is kept deformed under the constant shear strain $\gamma$. 
\end{enumerate}

To simulate quasi-static shear, we choose a sufficiently small $\dot{\gamma}= 10^{-4}$. We have checked that for $\varphi < \varphi_{\rm G}$, the stress-strain response is independent of $\dot{\gamma}$ when $\dot{\gamma}$ is decreased from $10^{-3}$ to $10^{-5}$ (see \modi{Supplementary Note 3}).

{\it Instantaneous shear --}
To simulate instantaneous shear, we instantaneously increase the shear strain from 0 to $\gamma$. We then turn on the harmonic soft-potential, and use the CG algorithm  to remove the overlaps. Different from the quasi-static shear, the system is generally far away from equilibrium after the instantaneous shear.

\bibliography{HSshear,HS,glass,Gardner,HS_yoshino,ref_yoshino}

\end{document}